\documentclass[11pt]{article}
\pdfoutput=1

\usepackage{jheppub}
\usepackage{subcaption}
\usepackage{amssymb}
\usepackage{import}
\usepackage{amsthm}
\usepackage{bigints}

\theoremstyle{remark}

\title{Amplituhedron meets Jeffrey-Kirwan Residue}
\author[1]{Livia Ferro,}\emailAdd{livia.ferro@lmu.de}
\author[2]{Tomasz \L ukowski,}\emailAdd{lukowski@maths.ox.ac.uk}
\author[2]{and Matteo Parisi}\emailAdd{parisi@maths.ox.ac.uk}

\affiliation[1]{Arnold--Sommerfeld--Center for Theoretical Physics,\\ Ludwig--Maximilians--Universit\"at, \\ Theresienstra\ss e 37, 80333 M\"unchen, Germany }
\affiliation[2]{Mathematical Institute, University of Oxford,\\ Andrew Wiles Building, Radcliffe Observatory Quarter,\\ Woodstock Road, Oxford, OX2 6GG, U.K.}

\abstract{The tree amplituhedra  $\mathcal{A}_{n,k}^{(m)}$ are mathematical objects generalising the notion of polytopes into the Grassmannian. Proposed for $m=4$ as a geometric construction encoding tree-level scattering amplitudes in planar $\mathcal{N}=4$ super Yang-Mills theory, they are mathematically interesting for any $m$. In this paper we strengthen the relation between scattering amplitudes and geometry by linking the amplituhedron to the Jeffrey-Kirwan residue, a powerful concept in symplectic and algebraic geometry. We focus on a particular class of amplituhedra in any dimension, namely cyclic polytopes, and their even-dimensional conjugates. We show how the Jeffrey-Kirwan residue prescription allows to extract the correct amplituhedron volume functions in all these cases. Notably, this also naturally exposes the rich combinatorial and geometric structures of amplituhedra, such as their regular triangulations.   
}

\begin{document}
\begin{flushright}
{\small LMU-ASC 27/18}
\end{flushright}
\maketitle

\addtocontents{toc}{\protect\setcounter{tocdepth}{1}}
\section{Introduction}
\label{introduction}

In the last decade, there have been remarkable developments in the realm of scattering amplitudes, mostly in planar $\mathcal{N}=4$ super Yang-Mills (SYM). In particular, we have seen lengthy expressions and involved computations from the era of Feynman diagrams replaced by many elegant mathematical structures. This was mostly due to a rephrasing of complicated algebraic and analytic problems in terms of geometry, which paved the way to new beautiful formulations and provided new insights for scattering amplitudes. 
In recent years, these culminated with the studies of positive Grassmannian \cite{ArkaniHamed:2012nw} and the discovery of the amplituhedron \cite{Arkani-Hamed:2013jha}.
The seed-idea was Andrew Hodges' geometrization of algebraic identities explaining the cancellation of spurious poles \cite{Hodges:2009hk}. This introduced the concept of ``scattering amplitudes as volumes of a geometric object", where different representations of the former are just different triangulations of the latter. 
Since its formulation, there have been intense studies of the amplituhedron \cite{Arkani-Hamed:2013kca,Franco:2014csa,Arkani-Hamed:2014dca,Ferro:2015grk,Galloni:2016iuj,Karp:2016uax,Ferro:2016zmx,Arkani-Hamed:2017vfh,Karp:2017ouj}. Nevertheless, many properties of this new mathematical object are still waiting to be discovered, see e.g.~\cite{Lam:2015uma} for some open questions.
Also recently, the paradigm of the amplituhedron has been further generalized into the one of positive geometries \cite{Arkani-Hamed:2017tmz}: these are spaces defined by positivity conditions that are always interwoven with a differential form which encodes physical quantities, e.g. scattering amplitudes. In addition to amplituhedra, examples of positive geometries relevant to physics include: the kinematic and worldsheet associahedra \cite{Arkani-Hamed:2017mur}, the Cayley polytopes \cite{He:2018pue} and the cosmological polytopes \cite{Arkani-Hamed:2017fdk}.
 
 In the spirit of positive geometries, the differential form of the amplituhedron is defined purely from the geometry by demanding it has logarithmic singularities on all the boundaries of the space. However, it is in practice often difficult to employ this geometric definition directly, and one needs to introduce spurious singularities which only cancel between various terms in the final answer. There have been several attempts to use the direct definition to express the differential form \cite{Arkani-Hamed:2014dca,Ferro:2015grk,Arkani-Hamed:2017tmz}, but most of the times one needs to refer to triangulations of the space. 
 Therefore, there is still a gap between the unified and beautiful geometric description of the amplituhedron and the algebraic-analytical nature of its differential form encoding the scattering amplitudes. In this paper we attempt to bridge this gap by introducing a new way to calculate the form based on the Jeffrey-Kirwan (JK) residue: our formula will encode all triangulations at once, in a way that is triangulation-independent. We remark the change of perspective compared to our previous work \cite{Ferro:2015grk}, where our aim was to obtain a formula which does not make any reference to triangulations.
 
The JK-residue was originally introduced by Jeffrey and Kirwan in their study of the localization of group actions \cite{JEFFREY1995291}: given a symplectic manifold and a group action on it, the residue formula relates elements of the equivariant cohomology of the manifold to the ones of the cohomology of its symplectic quotient. They were inspired by Witten \cite{Witten:1992xu} who re-examined the non-abelian localization of Duistermaat and Heckman \cite{Duistermaat1982} and applied it to physics obtaining new formulae in the context of two-dimensional Yang-Mills theory. 
Since then, the JK-residue has played an increasingly relevant role in physics, with particularly interesting applications in supersymmetric localization for gauge theories in various dimensions, e.g.~\cite{Benini:2013xpa,Closset:2015rna,Benini:2015noa}.
Most importantly for us, the notion of JK-residue has been also extended out of the realm of localization and can be generally regarded as an operation on rational differential forms, e.g. see \cite{BrionVergne,2004InMat.158..453S}. 

In the cases studied in this paper, the JK-residue applied to the amplituhedron provides the explicit expression for its logarithmic form and exposes the rich combinatorial structure of its triangulations.
In particular, when the amplituhedron is a cyclic polytope, the JK-residue naturally leads to the study of its secondary polytope \cite{MR1073208}, whose vertices correspond to the (regular) triangulations of the cyclic polytope. In this case, the fields charges familiar to aficionados of the Jeffrey-Kirwan prescription in supersymmetric localization are replaced by the Gale dual of a point configuration in  projective space. For positive external data, the Gale dual configuration of points provides a partition of the space into a set of chambers, each corresponding to a triangulation of the amplituhedron. By application of the JK-residue, they lead to distinct representations of the logarithmic differential form. 
Our method also extends to even-dimensional parity conjugate of cyclic polytopes, which are not polytopes any more. However, their triangulations are combinatorially equivalent to the ones of the corresponding cyclic polytope\footnote{After the completion of this paper we have become aware of \cite{Galashin:2018fri}, where the conjugation for the amplituhedron is also discussed.}. Finally, for odd-dimensional amplituhedra, the JK-residue prescription exposes a significant disparity between cyclic polytopes and their conjugates, casting doubt on a meaningful definition of conjugation in this case.

The paper is organized as follows: in Section 2 we recall the definition of the amplituhedron and its logarithmic differential form, with a special focus on the integral representation of the volume function. 
In Section 3 we introduce the JK-residue and discuss its properties. Section 4 illustrates the main statements of our paper and describes the geometric notions relevant for the cases discussed in the subsequent section. In Section 5 we present various examples of amplituhedra to explain thoroughly how our method applies for even- and odd-dimensional cyclic polytopes and their conjugates. Conclusions and open questions close the paper.


\section{The Amplituhedron}
In this section we collect some relevant information on the amplituhedron $\mathcal{A}_{n,k}^{(m)}$ that will be used in the remaining part of the paper. We start by recalling the original definition from \cite{Arkani-Hamed:2013jha}, which will allow us to set the notation we use in the subsequent sections. Then, following \cite{Ferro:2015grk}, we introduce the notion of volume function $\Omega_{n,k}^{(m)}$ and present its representation as an integral over a space of matrices. Finally, we give explicit results for $\Omega_{n,1}^{(m)}$ and introduce a novel representation for $\Omega_{n,n-m-1}^{(m)}$: they will be the two main objects of our study.
\subsection{Definition of the Amplituhedron}

The tree amplituhedron $\mathcal{A}_{n,k}^{(m)}$ is a generalization of the notion of polytopes into the Grassmannian. More precisely, it is a particular subset of the Grassmannian $G(m+k,k)$, which is the space of all $k$-planes $Y_{\alpha}^A$ in $(k+m)$ dimensions, $\alpha=1,\ldots,k$, $A=1,\ldots,k+m$. In order to define $\mathcal{A}_{n,k}^{(m)}$, we first fix positive external data given by $n$ vectors in $(k+m)$ dimensions: $Z_{i}^A$, $i=1,\ldots,n$, $A=1,\ldots,k+m$. Positivity means that all $(k+m)\times(k+m)$ ordered maximal minors of $Z$ satisfy 
\begin{equation}
\langle i_1,\ldots,i_{k+m} \rangle:=\langle Z_{i_1},\ldots,Z_{i_{k+m}}\rangle >0, \qquad \mathrm{for }\,\,i_1<\ldots<i_{k+m} \,,
\end{equation}
and therefore $Z$ is said to be an element of the positive Grassmannian $G_+(n,k+m)$.
Then the amplituhedron is the set of all $Y$ of the form
\begin{equation}\label{Ymap}
Y_{\alpha}^{A}=\sum_{i=1}^nc_{\alpha i}\,Z_i^A\,,
\end{equation}
for all matrices $C=(c_{\alpha i})_{\alpha=1,\ldots,k}^{i=1,\ldots,n}$ in the positive Grassmannian $G_{+}(n,k)$, i.e. for which all $k\times k$ ordered maximal minors of $C$ are positive
\begin{equation}
(i_1,\ldots,i_k):=(c_{i_1},\ldots,c_{i_k})>0\qquad\mathrm{for }\,\, i_1<\ldots <i_k\,.
\end{equation}

For the amplituhedron $\mathcal{A}_{n,k}^{(m)}$  there exists a canonical differential form $\mathbf{\Omega}_{n,k}^{(m)}$, which we refer to as the {\em volume form}, with the property that it has logarithmic singularities on all boundaries (of any dimension) of $\mathcal{A}_{n,k}^{(m)}$. 
Such differential form can always be written as
\begin{equation}
\mathbf{\Omega}_{n,k}^{(m)}=\prod_{\alpha=1}^k \langle Y_1\ldots Y_{k}d^mY_\alpha\rangle \Omega_{n,k}^{(m)}\,,
\end{equation}
 where $\Omega_{n,k}^{(m)}$ is a rational function which we refer to as {\em volume function}.
The main goal of this paper is to study the properties of this logarithmic form.

In particular, in the case $m=4$ the volume form encodes tree-level scattering amplitudes in planar $\mathcal{N}=4$ SYM. In this context, $n$ refers to the number of particles and $k$ the next-to-MHV degree. 
The variables $Z_i^A$ are the bosonized momentum twistors, i.e. a bosonized version of  momentum supertwistors \cite{Hodges:2009hk} $\mathcal{Z}_i^\mathcal{A}:=(\lambda_i^\alpha,\tilde\mu_i^{\dot\alpha},\chi_i^{\mathsf A}) $ with $\alpha,\dot\alpha=1,2$ and ${\mathsf A}=1,\ldots,4$. 
The tree-level amplitude is extracted as 
\begin{equation}\label{amplitoampli}
\mathcal{A}_{n,k}^{\mathrm{tree}}(\mathcal{Z})= \int d^{4\cdot k}\phi \,\, {\Omega}_{n,k}^{(4)}(Y^*,Z) \,,
\end{equation}
where we localized the volume form on an arbitrary $k$-dimensional reference plane $Y=Y^*$. The bosonized momentum twistor $Z_i$ projected on this plane is then expressed as $\phi^{\mathsf A}_\alpha\;\chi_{i\mathsf A}$ in terms of auxiliary Grassmann-odd variables $\phi_\alpha^\mathsf{A}$, while its four-dimensional projection on the complement of $Y^*$ is the momentum twistor $(\lambda_i^\alpha,\tilde\mu_i^{\dot\alpha})$. Without loss of generality, one can always choose  $Y^*=(\mathbb{O}_{k\times 4}\,|\,\mathbb{I}_{k\times k})$.

\subsection{Integral Representation of the Volume Function}

There are various ways to derive the logarithmic form $\mathbf{\Omega}_{n,k}^{(m)}$. In this paper we will focus on its integral representation: the corresponding volume function can be  expressed as an integral over a space of $k\times n$ matrices as
\begin{equation} \label{omegaintegral}
\Omega_{n,k}^{(m)}(Y,Z) = \int_{\gamma} \frac{d^{k \cdot n}\,c_{\alpha i}}{(1 2\ldots k) (2 3\ldots k+1)\ldots(n 1 \ldots k-1)} \prod_{\alpha=1}^k \delta^{m+k}(Y_{\alpha}^A - \sum_i c_{\alpha i}  Z_i^A)\,,
\end{equation}
where the integral is over a suitable contour $\gamma$, defined uniquely up to global residue theorems. Since the integrand is a rational function, such contour selects a set of poles and performing the integral reduces to evaluating an appropriate sum of residues. Each such residue corresponds to a particular $k\times m$-dimensional cell in the positive Grassmannian $G_+(n,k)$, whose image into the amplituhedron space through the map \eqref{Ymap} is top dimensional. We refer to such images as {\em generalized triangles}. The contour $\gamma$ in \eqref{omegaintegral} is such that the corresponding union of generalized triangles provides a triangulation of the amplituhedron  $\mathcal{A}_{n,k}^{(m)}$.

While, in general, the correct contour computing the volume function $\Omega_{n,k}^{(m)}$ is not known \emph{a priori}, for the $k=1$ case an appropriate prescription (called ``$i\epsilon$'') was proposed in \cite{Nima:2014,Ferro:2015grk,Arkani-Hamed:2017tmz}.
 The procedure described there is iterative and requires a careful study of relations in the space of $\epsilon$-parameters, which makes it technically involved. In this paper we propose a simpler, though more powerful method, which also provides interesting insights into geometric properties of the amplituhedron and its triangulations.
It is based on the Jeffrey-Kirwan residue which, together with many applications in various branches of mathematics, plays a central role in supersymmetric localization.
In this paper we systematically developed our observation in two cases: $k=1$ for any $m$ and $n-m-k=1$ for even $m$.


\subsection{Volume Function for \texorpdfstring{$\mathbf{k=1}$}{}, all \texorpdfstring{$\mathbf{m}$}{}}

 In the $k=1$ case, the amplituhedron is a cyclic polytope with $n$ vertices in $m$ dimensions, denoted by $\mathcal{C}(n,m)$, and the integral \eqref{omegaintegral} reduces to
\begin{equation} \label{omegaintegralk1}
\Omega_{n,1}^{(m)}(Y,Z) = \int_{\gamma} \frac{d^{  n}\,c_{i}}{c_1\ldots c_n} \delta^{m+1}(Y^A - \sum_i c_{ i}  Z^A_i)\,.
\end{equation}
This class of polytopes has been extensively studied over the years: for a comprehensive introduction to their interesting properties we refer the reader to the book \cite{de2010triangulations} and references therein. For example, cyclic polytopes have the largest number of faces of every dimension among all polytopes of a fixed dimension and number of vertices; they also have many triangulations. In this paper we show how the structure of their (regular) triangulations is captured by the integral \eqref{omegaintegralk1} and the corresponding Jeffrey-Kirwan residue. 

For $k=1$ and any value of $m$, generalized triangles are simplices defined as convex hulls of points $\{Z_{i_1},\ldots,Z_{i_{m+1}}\}$. Their volume functions have very simple representations in terms of the brackets
\begin{equation}\label{brackets}
[i_1\ldots i_{m+1}]=\frac{\langle i_1\ldots i_{m+1}\rangle^m}{\langle Y i_1 i_2 \ldots i_{m}\rangle\langle Y i_2 i_3 \ldots i_{m+1}\rangle\ldots \langle Y i_{m+1}i_1 \ldots i_{m-1}\rangle}\,.
\end{equation}
Each such bracket can also be interpreted as the actual volume of the simplex dual to the  respective generalized triangle.
Using this notation we can write the volume function $\Omega_{n,1}^{(m)}$ for all values of $m$, see \cite{Arkani-Hamed:2017tmz}. We list below the results relevant for the examples we will study in the next sections  
\begin{align}\label{answer2}
\Omega_{n,1}^{(1)}&=\sum_{i}^{}\, [i\,i+1]\,, \qquad\qquad\qquad\Omega_{n,1}^{(2)}=\sum_{i}^{}\, [1\,i\,i+1]\,,\\\label{answer4}
\Omega_{n,1}^{(3)}&=\sum_{i<j}\, [i\,i+1\,j\,j+1]\,,\qquad\,\,\,\,\,\Omega_{n,1}^{(4)}=\sum_{i<j}\, [1\,i\,i+1\,j\,j+1]\,.
\end{align}

\subsection{Volume Function for \texorpdfstring{$\mathbf{n-m-k=1}$}{}, even \texorpdfstring{$\mathbf{m}$}{}}
\label{conjugateampl}
In the context of scattering amplitudes in $\mathcal{N}=4$ SYM, there is a natural operation, parity conjugation, which maps N$^k$MHV amplitudes into their $\overline{\mathrm{N}^k\mathrm{MHV}}$ counterparts. In the amplituhedron description, this conjugation can be generalized for any $m$ and it is realized by replacing $k$ with $n-m-k$, see for example the discussion in \cite{Arkani-Hamed:2017vfh}. The conjugate amplituhedron $\overline{\mathcal{A}}_{n,k}^{(m)}=\mathcal{A}_{n,n-m-k}^{(m)}$ is then the set of point in $G(n-k,n-k-m)$ such that
\begin{equation}
\overline{Y}_{\bar\alpha}^{\bar A}=\sum_{i=1}^n\overline{c}_{\bar\alpha\,i}\overline{Z}_i^{\bar A}\,,\qquad \overline{c}\in G_+(n,n-k-m)\,,\quad\overline{Z}\in G_+(n,n-k)\,,
\end{equation}
with $\bar\alpha=1,\ldots,n-k-m$ and $\bar A=1,\ldots,n-k$.
Interestingly, many combinatorial properties of the conjugate amplituhedron $\overline{\mathcal{A}}_{n,k}^{(m)}$ follow closely the ones of $\mathcal{A}_{n,k}^{(m)}$. As an example, we observed that the number of all generalized triangles is the same in the two cases. This statement is very non-trivial since the two amplituhedra live in spaces with different number of dimensions. However, it does not imply that the amplituhedron and its conjugate can be mapped into each other. Indeed, when $m$ is odd we find that the combinatorial structure of their triangulations is inequivalent. Nevertheless, for even $m$ and at least for $k=1$, one can also find a map which relates all triangulations of the amplituhedron and its conjugate, making them indistinguishable from the combinatorial point of view.

In order to outline our claim,  we present here a novel representation of the amplituhedron volume function for even $m$ and $n-k-m=1$, in a form which makes the relation with its conjugate case, $k=1$, manifest. In particular, we observe that in both cases the volume function can be written exactly in the same form in terms of the brackets \eqref{brackets} or their conjugate counterparts which we define in the following. We start by introducing some relevant notation. Let us denote by
\begin{equation}
\underline{i}=(i,i+1,\ldots,i+k-1)\,,
\end{equation}
subsets of consecutive indices, $i=1,\ldots,n$. All indices are understood modulo $n$. Building blocks for the $n-k-m=1$ volume functions are brackets labelled by $m+1$ such indices. For any set $\{\underline{i}_1,\ldots,\underline{i}_{m+1}\}$ let us define $\{\underline{j}_1,\ldots,\underline{j}_{k}\}$ such that $\{{i}_1,\ldots,{i}_{m+1},{j}_1,\ldots,{j}_{k}\}=\{1,\ldots,n\}$ as a set. Then we define conjugate brackets as
\begin{equation}\label{conjugatebracket}
\overline{[{i}_1\ldots{i}_{m+1}]}=\frac{\left(\det\limits_{\alpha,\beta} \langle Y^{k-1}_\alpha \underline{\overline{j}}_{\beta}\rangle\right)^m}{[Y;\underline{i}_1]\ldots [Y;\underline{i}_{m+1}]}\,,
\end{equation}
where
\begin{equation}
[Y;(l_1,\ldots,l_k)]=\left(\mathrm{sign}(j_1,\ldots,j_k,l_1)\right)^{k+1}\,\det\limits_{\alpha,\beta} \langle Y \overline{\underline{j}_{\alpha}\cup l_\beta}\rangle\,.
\end{equation}
If $l_\beta\in \underline{j}_\alpha$, then we set $\langle Y \overline{\underline{j}_{\alpha}\cup l_\beta}\rangle=0$. Here, the overline indicates a complementary subset of indices inside the set $\{1,\ldots,n\}$ and $\mathrm{sign}(j_1,\ldots,j_k,l_1)$ is the sign of the permutation.

Since the definition \eqref{conjugatebracket} is quite involved, we present here a simple explicit example for $m=2$. Let us calculate $\overline{[1,2,4]}$. In this case $\underline{i}_1=(1,2),\underline{i}_2=(2,3),\underline{i}_3=(4,5)$ and  $\underline{j}_1=(3,4),\underline{j}_2=(5,1)$. Then $\overline{\underline{j}_1}=(1,2,5),\overline{\underline{j}_2}=(2,3,4)$ and the numerator becomes
\begin{equation}
(\det\limits_{\alpha,\beta} \langle Y^{k-1}_\alpha \underline{\overline{j}}_{\beta}\rangle)^2  = (\langle Y_1 125\rangle \langle Y_2 234\rangle-\langle Y_2 125\rangle \langle Y_1 234\rangle)^2 =:\langle Y(125)\cap (234)\rangle^2\,.
\end{equation}
The denominators are:
\begin{align}
[Y;\underline{i}_1]&=[Y;(1,2)]=\left| \begin{tabular}{cc}
$\langle Y 25 \rangle$&$\langle Y 15\rangle$\\
0&$\langle Y34 \rangle$
\end{tabular}\right|=\langle Y25 \rangle \langle Y34\rangle\,,\\
[Y;\underline{i}_2]&=[Y;(2,3)]=\left| \begin{tabular}{cc}
$\langle Y 15 \rangle$&0\\
$\langle Y34 \rangle$&$\langle Y24 \rangle$
\end{tabular}\right|=\langle Y15 \rangle \langle Y24\rangle\,,\\
[Y;\underline{i}_3]&=[Y;(4,5)]=-\left| \begin{tabular}{cc}
0&$\langle Y 12 \rangle$\\
$\langle Y 23\rangle$&0
\end{tabular}\right|=\langle Y12 \rangle \langle Y23\rangle\,.
\end{align}
Collecting all terms we get
\begin{equation}
\overline{[124]}=\frac{\langle Y(125)\cap (234)\rangle^2}{\langle Y 12\rangle\langle Y 25\rangle\langle Y 51\rangle\langle Y 23\rangle\langle Y 34\rangle\langle Y 42\rangle}\,,
\end{equation}
which we can recognize as the usual representation of building blocks for the $m=2$, $k=2$ case, see \cite{Arkani-Hamed:2017tmz}. For $m=2$, the terms in the denominator of \eqref{conjugatebracket} always factorize. This is however not true for $m>2$.

Using the brackets \eqref{conjugatebracket}, the volume  function for $m=2$ can be written as
\begin{equation}
\label{volumedualm2}
\Omega_{n,n-3}^{(2)}=\sum_{i}^{}\, \overline{[1\,i\,i+1]}\,,
\end{equation}
while for $m=4$ it is
\begin{equation}
\Omega_{n,n-5}^{(4)}=\sum_{i<j}^{}\, \overline{[1\,i\,i+1\,j\,j+1]}\,,
\end{equation}
making it manifestly conjugate to the $k=1$ case.


\section{Jeffrey-Kirwan Residue}
\label{Sec:JKresidue}

 Our exposition of the Jeffrey-Kirwan residue method follows \cite{Benini:2013xpa} and it is tailored for the subsequent application to the calculation of the amplituhedron volume functions.  

Consider a real vector space $V=\mathbb{R}^r$ and its elements $x=(x_1,\ldots,x_r) \in V$.
Let $\omega$ be a rational differential form on $V$
\begin{equation}\label{diff.form}
\omega = \frac{dx_1 \wedge \ldots \wedge dx_r}{\beta_1(x) \ldots \beta_n(x)} \,,
\end{equation}
where $r\leq n$ and all denominators are affine-linear functions
\begin{equation}\label{betas.x}
\beta_i(x) = \beta_i \cdot x + \alpha_i \,, \quad  \beta_i \in V^*, \, \alpha_i \in \mathbb{R}\,,\quad i=1,\ldots,n\,.
\end{equation}
 In the following, we refer to $\beta_i$ as {\em charges} 
and define their collection $B=\{\beta_i\}_{i=1,\ldots,n}$ which is a discrete subset in the dual space $V^*=\mathbb{R}^r$. 
We assume that $\beta_i(x)$ are in {\em general position} which, by definition, means that 
\begin{align}
\bigcap_{j=1}^{r}\{\beta_{i_j}(x)=0\}=\{point\} \,,\qquad\qquad \bigcap_{j=1}^{r+1}\{\beta_{i_j}(x)=0\}=\emptyset\,,
\end{align}
for any subset $\{i_1,\ldots,i_r,i_{r+1}\}\subseteq \{1,\ldots,n\}$. This condition ensures that all poles of the differential form \eqref{diff.form} are discrete points and simple.

Let us consider a subset ${A}$ of ${B}$. We call ${A}$ a {\em basis} if it forms a basis of $\mathbb{R}^r$ and we denote the set of all bases by $\mathcal{B}(B)$. For each ${A} = \{\beta^A_{1},\ldots,\beta^A_{r}\} \, \in \, \mathcal{B}(B)$ we define a {\em cone}
\begin{equation}\label{cones}
\mathrm{Cone}_{A}:=\left\{\sum_{j=1}^r a_j \beta^A_{j} \in \mathbb{R}^r \rvert a_1,\ldots,a_r > 0\right\} \,,
\end{equation}
as the convex hull of $r$ vectors in $A$.
Since $A$ is a basis, the cone is full-dimensional and simplicial. 
To each basis $A\in\mathcal{B}(B)$ corresponds a pole $x_A$ of the rational form $\omega$ defined by the solution to $\{\beta^A_{1}(x_{{A}})=0,\ldots,\beta^A_{r}(x_{{A}})=0\}$. We assumed that $\beta_i(x)$ are in general position therefore $x_{A}$ is unique and  $\beta_i(x_A)\neq0$ for all $\beta_i\not\in A$. Finally, we say that $\eta\in V^*$ is {\em generic with respect to $B$} if $\eta$ does not belong to the boundary of any of the cones $\mathrm{Cone}_A$ for all $A\in \mathcal{B}(B)$.

We are now ready to define the Jeffrey-Kirwan residue of $\omega$ with respect to the set of charges $B$ and a generic vector $\eta\in V^*$ 
\begin{equation}
\mathrm{JKRes}^{B,\eta} \, \omega  =  \sum_{A \in \mathcal{B}(B)} \underset{x=x_{{A}}}{\mathrm{JKRes}}^{B,\eta} \omega\,,
\end{equation}
where the sum runs over all bases of $B$ and for a given basis we have
\begin{equation}
\underset{x=x_{{A}}}{\mathrm{JKRes}}^{B,\eta} \omega  = 
                \begin{cases}
               \frac{1}{|\mathrm{det}(\beta^A_{1} \ldots \beta^A_{r})|} \frac{1}{\prod_{\beta_j \notin {A}}\beta_j(x_{{A}})} \,, & \eta \in \mathrm{Cone}_{{A}}  \\
                  0\,,  &\mathrm{otherwise}
                \end{cases}\,.
\end{equation}
We emphasize that, in the case when $\eta\in \mathrm{Cone}_A$, this definition agrees with the standard multivariate residue of $\omega$ at the point $x_A$, up to a sign determined by the orientation of charges defining  $\mathrm{Cone}_A$.

We end this section with the definition of chamber, which will be useful later on to relate the Jeffrey-Kirwan residue to amplituhedron triangulations. It was shown in \cite{JEFFREY1995291} that the JK-residue $\mathrm{JKRes}^{B,\eta}$ is independent of  $\eta$. Various choices of $\eta$ might however lead to a different form of the answer.  Therefore, we would like to find equivalence classes of vectors $\eta\in V^*$ which give exactly the same form of the answer. Such equivalence classes are precisely the connected components of the set of all generic vectors $\eta\in V^*$  which we call {\em chambers}. We introduce two characterizations of chambers. First, two vectors are contained in the same chamber if the set of cones to which they belong is the same. More precisely, if we define $\mathrm{Cone}(\eta) = \{A\in \mathcal{B}(B) : \eta \in \mathrm{Cone}_A\}$ then a chamber including a generic vector $\eta_*$ is the set $ \{\eta \in \mathbb{R}^r : \mathrm{Cone}(\eta) = \mathrm{Cone}(\eta_*)\}$. The second characterization is given by maximal intersections of cones and provides an algorithmic procedure to find all chambers. Let us define the set of all cones and their non-trivial intersections
\begin{equation}\label{poset}
\Lambda=\{\mathrm{Cone}_{A_{i_1},\ldots,A_{i_p}}:=\mathrm{Cone}_{A_{i_1}}\cap\ldots\cap \mathrm{Cone}_{A_{i_p}}\neq \emptyset: A_{i_j}\in \mathcal{B}(B)\}\,.
\end{equation} 
We define a partial order $\prec$ on $\Lambda$ given by the inclusion
\begin{equation}\label{partial.order}
\mathrm{Cone}_{A_{i_1},\ldots,A_{i_p}}\prec\mathrm{Cone}_{A_{j_1},\ldots,A_{j_s}} \,\,\mathrm{iff} \,\,\mathrm{Cone}_{A_{i_1},\ldots,A_{i_p}}\subset\mathrm{Cone}_{A_{j_1},\ldots,A_{j_s}}\,.
\end{equation}
Then chambers are minimal elements in the ordering $\prec$. Contrary to the simplicial cones defined in \eqref{cones}, chambers are often polyhedral, which means that they are convex hulls of $r'$ vectors  with $r'>r$.
\section{Amplituhedron from Jeffrey-Kirwan Residue}
\subsection{Main Statement}\label{Sec:mainstatement}

Let us present here the main statement of this paper that provides a way to extract the amplituhedron volume functions from the integral \eqref{omegaintegral} using the JK-residue. We start from the $k=1$ case and consider the integral \eqref{omegaintegralk1}, which is an $n$-dimensional integral with $m+1$ delta functions. After we solve the delta functions, the integrand becomes an $(n-m-1)$-dimensional rational differential form with exactly $n$ denominators in $n-m-1$ variables. Without loss of generality, we can keep the first $n-m-1$ variables $x=(c_1,\ldots,c_{n-m-1})$ and solve
\begin{equation}\label{YiscZ}
Y^A=\sum_{i=1}^nc_i \,Z^A_i\,,\qquad A=1,\ldots,m+1\,,
\end{equation}
for the remaining ones
\begin{equation}
c^*_i(x)\,,\qquad i=n-m,\ldots,n\,.
\end{equation}
Then the integrand becomes
\begin{equation}\label{omegaexpl}
\omega_{n,1}^{(m)}=\frac{dx_1 \wedge \ldots \wedge dx_{n-m-1}}{x_1\ldots x_{n-m-1} \,c^*_{n-m}(x)\ldots c^*_{n}(x)} \,,
\end{equation}
where the functions $\beta_i(x)$ in \eqref{diff.form} have the explicit form
\begin{equation}\label{cstars}
\beta_i(x)=\begin{cases}x_i&i=1,\ldots,n-m-1\\c_i^*(x)&i=n-m,\ldots,n\end{cases}\,.
\end{equation}
Our main claim is that, if the external data $Z_i^A$ is positive, then the volume function of the amplituhedron can be obtained from the Jeffrey-Kirwan residue as
\begin{equation}\label{omegafromJK}
\Omega_{n,1}^{(m)}=\mathrm{JKRes}^{B,\eta}\,\omega_{n,1}^{(m)}\,,
\end{equation}
where $B=\{\beta_i\}_{i=1,\ldots,n}$ and the charges $\beta_i$ are defined from \eqref{cstars} using \eqref{betas.x}.
Here $\eta\in\mathbb{R}^{n-m-1}$ is any generic element with respect to $B$. For all $\eta$ in a given chamber the form of the answer we get from \eqref{omegafromJK} is the same, whereas answers obtained in various chambers are different representations of the same function. They correspond to distinguished choices of the contour $\gamma$ in \eqref{omegaintegralk1}, related to each other by global residue theorem. From the geometric point of view, different contours correspond to different triangulations of the amplituhedron. This establishes the correspondence between chambers and triangulations, which is explained in greater details in the following sections.

We now consider the case of $n-k-m=1$ and even $m$:
\begin{equation} \label{omegaintegralconjugate}
\Omega_{n,n-m-1}^{(m)}(Y,Z) = \int_{\gamma} \frac{d^{k\times n}c}{c_{\underline{1}}\ldots c_{\underline{n}}} \prod_{\alpha=1}^{k}\delta^{n-1}(Y^A_\alpha - \sum_i c_{\alpha i}  Z^A_i)\,,
\end{equation}
where we introduced the notation $c_{\underline{i}}=(i\,i+1\ldots i+k-1)^{\mathrm{ord}}$ and $()^\mathrm{ord}$ indicates that the columns of the minor are ordered\footnote{For example for $n=6$ and $k=3$, we have $c_{\underline{5}}=(156)$.}. This notation exposes the fact that this case is conjugate to $k=1$. After we solve the delta functions in \eqref{omegaintegralconjugate} and change variables, the integrand becomes the following rational form
\begin{equation}\label{omegaconjugate}
\omega_{n,n-m-1}^{(m)}=\frac{dx_{1} \wedge \ldots \wedge dx_{k}}{x_{1}\ldots x_{k} \,c^*_{\underline{k+1}}(x)\ldots c^*_{\underline{n}}(x)} \,.
\end{equation}
We observe that $c^*_{\underline{i}}(x)$ are linear functions of $x$: analogously to \eqref{cstars}, we can read off the functions $\beta_{\underline{i}}(x)$ and find the set of charges $\overline{B}$. For positive external data, this leads us to the following claim, similar to \eqref{omegafromJK}: 
\begin{equation}\label{omegaconjugatefromJK}
\Omega_{n,n-m-1}^{(m)}=\mathrm{JKRes}^{\overline{B},\eta}\,\omega_{n,n-m-1}^{(m)}\,,
\end{equation}
where, as before, we pick a generic vector $\eta\in \mathbb{R}^{n-m-1}$, and the answer is independent of it. Choosing $\eta$ in different chambers is related to different triangulations of the amplituhedron $\mathcal{A}_{n,n-m-1}^{(m)}$ and therefore to different representations of the volume function $\Omega_{n,n-m-1}^{(m)}$.


\subsection{Gale Transform}
For $k=1$, in order to construct the rational differential form $\omega_{n,1}^{(m)}$ in \eqref{omegaexpl}, we have solved the constraints \eqref{YiscZ} and obtained the set of charges $B$. Notice that they do not depend on $Y$ but purely on the external data $Z$. In this section we argue that there exists an alternative perspective on how to interpret the charges $\beta_i$: they are obtained via a procedure called {\em association} or {\em Gale transform} of a configuration of points.

In order to define the Gale transform, we consider a generic ordered collection ${\bf w}=(w_1,\ldots,w_n)$ of $n$ points in a $d$-dimensional space, $d<n$, and arrange their coordinates to form a $d\times n$ matrix. The row-space of such matrix defines a $d$-dimensional subspace $W$ of the $n$-dimensional space. Let us take its orthogonal complement  $W^\perp$ which can be represented as an $(n-d)\times n$ dimensional matrix. It is defined up to a $GL(n-d)$ redundancy which we fix by setting the first $n-d$ columns to be the identity. The column vectors of this matrix define a collection ${\bf w^\perp}=(w_1^\perp,\ldots,w_n^\perp)$ of $n$ points in an $(n-d)$-dimensional space. The configuration ${\bf w^\perp}$ is by definition the Gale dual to ${\bf w}$. In the light of this definition, it is obvious that the Gale dual of the amplituhedron external data $Z_i^A$ is exactly the set of charges we defined in \eqref{cstars}.

One can find a general, explicit form of Gale transform for any configuration of $n$ points in $d=m+1$ dimensions and it is given by
\begin{equation}\label{Zperp}
(Z_i^\perp)^{\dot{A}}=\begin{cases}\delta_{i}^{\dot{A}}\,,&i=1\ldots,,n-m-1\\-\frac{\langle n-m,\ldots,n\rangle|_{i\to \dot{A}}}{\langle n-m,\ldots,n\rangle}\,,& i=n-m,\ldots,n \end{cases}\,,
\end{equation}
where $\dot A=1,\ldots,n-m-1$. Here, $\langle\rangle|_{i\to \dot A}$ indicates that we replace the vector $Z_i$ by $Z_{\dot A}$ inside the bracket.
In the amplituhedron context the external data is always positive, i.e.~the matrix $Z_{i}^A$ is an element of the positive Grassmannian, which implies that the polytope it defines is convex. Directly from \eqref{Zperp} we see that the Gale dual configuration is an element of the Grassmannian $G(n,n-m-1)$; it does not belong, however, to its non-negative part. Nevertheless, one can show that $0$ is inside the convex hull of $Z_i^\perp$, $i=1,\ldots,n$ \cite{BILLERA1990155}. This has the important geometric implication that the union of all closures of cones defined by the Gale dual configuration covers the whole dual space, i.e.
\begin{equation}\label{fulcover}
\bigcup_{A\in \mathcal{B}(B)}\overline{\mathrm{Cone}_A}=\mathbb{R}^{n-d}\,.
\end{equation}   
If this was not the case, there would exist a generic vector not included in any cone and therefore the JK-residue would be identically zero.

Until now, we have only discussed the $k=1$ case. It would be advantageous to make a similar statement about the Gale transform also for the conjugate amplituhedron when $n-m-k=1$. Notice, however, that in this case the charges depend explicitly also on $Y$. For this reason it is unclear how to obtain them from the Gale transform. Therefore, solving the $\delta$-functions and extracting the proper coefficients of the linear factors in the denominator of \eqref{omegaconjugate} remains the only way to find charges when $n-m-k=1$.

\subsection{Secondary Polytope}
The Gale transform is a very powerful method for the study of triangulations of polytopes. We review definitions and results, mainly following \cite{gelfand2008discriminants}, in order to explain this approach.

A \emph{triangulation} of a convex polytope $\mathcal{P}$ is a set of simplices which together cover $\mathcal{P}$ and are intersecting properly, i.e. their pair-wise intersection is either a shared boundary or an empty set.
For the purpose of the paper, we describe a particular class of triangulations, which can be constructed as follows.
To each vertex $P_i$ of the polytope $\mathcal{P}$ we assign a weight $\alpha(P_i) \in \mathbb{R}$. Then, the convex hull of the half-lines $\{(P_i,z),  z \leq \alpha(P_i) \} \subset \mathbb{R}^{d}\times \mathbb{R}$ is an unbounded polyhedron with vertices $\{(P_i,\alpha(P_i))\}$. It can be shown that, if the weights are generic enough, the bounded faces of such a polyhedron are simplices, and their projections to $\mathbb{R}^d$ form a triangulation of the polytope $\mathcal{P}$. All triangulations of $\mathcal{P}$ arising in this way are called \emph{regular}.
It is interesting to notice that, even for well-behaved objects like cyclic polytopes, there are in general more non-regular triangulations than regular ones, with the former appearing already in three dimensions for the case of nine vertices \cite{ATHANASIADIS200019}. The set of regular triangulations is known to have a friendly structure. In particular, one can show \cite{gelfand2008discriminants} that there exists a polytope in $n-d-1$ dimensions\footnote{The original construction realizes $\Sigma(\mathcal{P})$ as the convex hull of the so-called \emph{GKZ-vectors} in $n$ dimensions and then shows that they indeed lie on a $(n-d-1)$-plane.}, the \emph{secondary polytope} $\Sigma(\mathcal{P})$ of $\mathcal{P}$, whose vertices are in one-to-one correspondence with regular triangulations of $\mathcal{P}$. Moreover, the face poset of $\Sigma(\mathcal{P})$ is isomorphic to the poset of all regular subdivisions of $\mathcal{P}$. In particular, the edges of a secondary polytope are connecting triangulations related by the so-called bistellar flips that, as we will show in the following sections, correspond to global residue theorems for the integral \eqref{omegaintegralk1}.  

In order to make a connection with the previous sections, it is now useful to introduce the notion of  \emph{fan}: a set of  cones in a vector space $V$ which are properly intersecting and together cover $V$ fully.
If $\mathcal{P}$ is a polytope and $P_i$ one of its vertex, we define the \emph{normal cone} of $\mathcal{P}$ at the point $P_i$ to be the cone generated by the vectors normal to all faces of $\mathcal{P}$ containing $P_i$. 
We can introduce the \emph{normal fan} of $\mathcal{P}$ as the collection of the normal cones of $\mathcal{P}$ at all of its vertices. 
Finally, we define the \emph{secondary fan}\footnote{In \cite{gelfand2008discriminants} the notion of \emph{secondary fan} is introduced in a different fashion and subsequently proved to be the normal fan of the secondary polytope.} of a polytope $\mathcal{P}$ as the normal fan of its secondary polytope $\Sigma(\mathcal{P})$.

We are now ready to draw the connection between space of charges and triangulations.
Consider a polytope $\mathcal{P}$ in the projective space $\mathbb{P}^{d}$ with vertices ${\bf w}=(w_1,\ldots,w_n)$ written using $(d+1)$-dimensional  homogeneous coordinates. We explained above how to perform the Gale transform of these points to obtain ${\bf w^\perp}=(w_1^\perp,\ldots,w_n^\perp)$ in the $(n-d-1)$-dimensional space of charges. Let us study the chambers corresponding to this configuration of points. By their characterization as maximal intersection of cones \eqref{poset}, \eqref{partial.order} and the property \eqref{fulcover}, it is easy to see that the collection of all chambers is itself a fan, called the \emph{chamber fan}.
The main statement of this section is that the chamber fan of $\mathcal{P}$ is precisely the secondary fan of the polytope. In simpler words, if for each chamber we pick a unit vector contained in it and we consider the convex hull of all such vectors, we obtain a polytope which has a combinatorial structure equivalent to the secondary polytope.
This establishes the exact correspondence between regular triangulations of the amplituhedron for $k=1$, as vertices of the corresponding secondary polytope, and the chambers in the space of charges $\mathbb{R}^{n-m-1}$.

\subsection{Global Residue Theorem}
In order to give an interpretation to the edges of a secondary polytope, we recall some details on the global residue theorem for multivariate functions. 
Let us study again the rational differential form 
\begin{equation}
\omega = \frac{dx_1 \wedge \ldots \wedge dx_r}{\beta_1(x) \ldots \beta_n(x)} \,.
\end{equation}
For $r<n$ we use the global residue theorem as follows. Let us partition the denominators into $r$ divisors $D_i(x)$ such that
\begin{equation}
\prod_{i=1}^r D_{i}(x)=\prod_{j=1}^n \beta_{j}(x)\,.
\end{equation}
Since $\beta_i(x)$ are in general position then the intersection $S=\{D_{1}(x)=0\}\cap\ldots\cap\{ D_{r}(x)=0\}$ consists of a finite number of discrete points. Then, the global residue theorem reads
\begin{equation}\label{GRT.general}
\sum_{P\in S} \mathrm{Res}_P \,\omega=0\,.
\end{equation}
In the following we will see that the Jeffrey-Kirwan residue provides a purely geometric realization of these relations. Indeed, if we consider two different vectors $\eta$ and $\tilde\eta$ belonging to two different chambers, we can find a finite sequence of global residue theorems which relate $\mathrm{JKRes}^{B,\eta}\,\omega_{n,1}^{(m)}$ to $\mathrm{JKRes}^{B,\tilde\eta}\,\omega_{n,1}^{(m)}$. In particular, the global residue theorem relating two adjacent chambers, and therefore two adjacent vertices of the secondary polytope, corresponds to a bistellar flip, whose geometric origin will be explained in the next section. 

\section{Examples}
\subsection{Preliminary Remarks}

In this section we give a detailed treatment of the notions introduced in this paper providing some simple examples. Many geometric aspects for the $k=1$ case have been already documented in the literature, see \cite{de2010triangulations}. We explain how they are reproduced by applying the JK-residue method to the integral \eqref{omegaintegral}, deriving all possible representations of the amplituhedron volume form related to regular triangulations. 
Moreover, we show that the previously unknown structure of triangulations for the $n-k-m=1$ case for even $m$ is identical to the one of cyclic polytopes, confirming they are objects conjugate to each other.
We start our analysis from the case $m=2$, which leads to the study of $n$-gons and their triangulations described by the associahedron. Then we move to the case relevant for physics, namely $m=4$. Finally, we discuss odd-dimensional cyclic polytopes. We end this section with comments on the case $n-m-k=1$. 

For each $k=1$ example, we start by finding the set of charges $B$ using the Gale transform of external data $Z_i^A$. As the second step, we get all chambers by finding minimal elements in the partial order $\prec$ defined in \eqref{partial.order}. For each chamber we pick a vector $\eta$ belonging to it and calculate the Jeffrey-Kirwan residue  $\mathrm{JKRes}^{B,\eta}\,\omega_{n,1}^{(m)}$. The answer for each chamber can be written in terms of the brackets \eqref{brackets}. For $n-k-m=1$ we follow the same steps, apart from the derivation of charges, which are obtained by directly solving the $\delta$-functions instead of using the Gale transform. We write the final answer for volume forms using the brackets~\eqref{conjugatebracket}.

\subsection{Toy Amplituhedron: \texorpdfstring{$\mathbf{m=2}$}{}}
\label{Sec:toy}
\begin{figure}[ht]
\centering{
\def\svgwidth{0.5\linewidth}{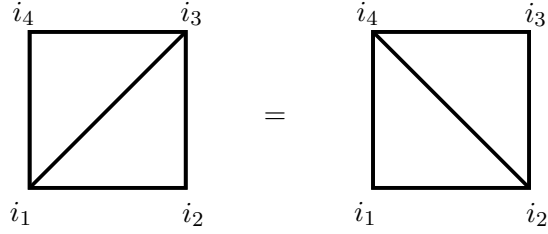}
}
\caption{Global residue theorem or bistellar flip for $m=2$.}
\label{fig:GRTm2}
\end{figure}

Let us start by studying the toy example of the tree amplituhedron, i.e. $m=2$, which will allow us to explore the notions we introduced in previous sections, making it easy to visualize them. In this case the external data defines a convex polygon in the projective space $\mathbb{P}^2$. All triangulations of any $n$-gon are classified and they are known to be vertices of the so-called associahedron $K_{n-1}$. This is an $(n-3)$-dimensional convex polytope with the number of vertices given by the Catalan number $C_{n-2}=\frac{1}{n-1}\binom{2n-4}{n-2}$. As we mentioned before, each triangulation of the $n$-gon corresponds to a chamber in the space of charges. Every chamber produces a different sum of cones in the JK-residue computation, giving a distinguished representation of the volume function for each triangulation. These representations are related to each other by global residue theorems, which in the case of $m=2$ are generated by the simple four-terms identity
\begin{equation}\label{GRTm2}
[i_1i_2i_3]+[i_1 i_3i_4]=[i_1i_2i_4]+[i_2i_3i_4]\,, \qquad 1\leq i_1<i_2<i_3<i_4\leq n\,,
\end{equation} 
which is depicted in Fig.~\ref{fig:GRTm2}. This formula can be derived by using the global residue theorem \eqref{GRT.general} with
\begin{equation}
D_l(x)=\beta_{j_l}(x)\,,\quad l=1,\ldots,n-4\,, \quad\mbox{and}\quad
D_{n-3}(x)=\beta_{i_1}(x)\beta_{i_2}(x)\beta_{i_3}(x)\beta_{i_4}(x)\,,
\end{equation}
where $j_1,\ldots,j_{n-4}$ are complementary indices to $i_1,\ldots,i_4$.  From the geometric point of view, this relation represents two ways of triangulating a square by flipping its diagonal. Such elementary transformation connecting two triangulations is usually referred to as the (two-dimensional) {\em bistellar flip}. In order to relate two arbitrary representations of the volume form we need to use a sequence of bistellar flips, where at each step they connect two adjacent chambers, namely two chambers sharing codimension-one boundaries. Moreover, it is important to notice that two triangulations related by a bistellar flip have the same number of triangles. From the point of view of the JK-residue, this is manifested by the fact that each chamber is contained in the same number of cones, which for a given $n$ is exactly $n-2$. 

\bigskip
\noindent {\bf Example: $\mathbf{n=4}$.}

\begin{figure}[ht]
\centering{
\begin{subfigure}{0.4\textwidth}
\def\svgwidth{\linewidth}{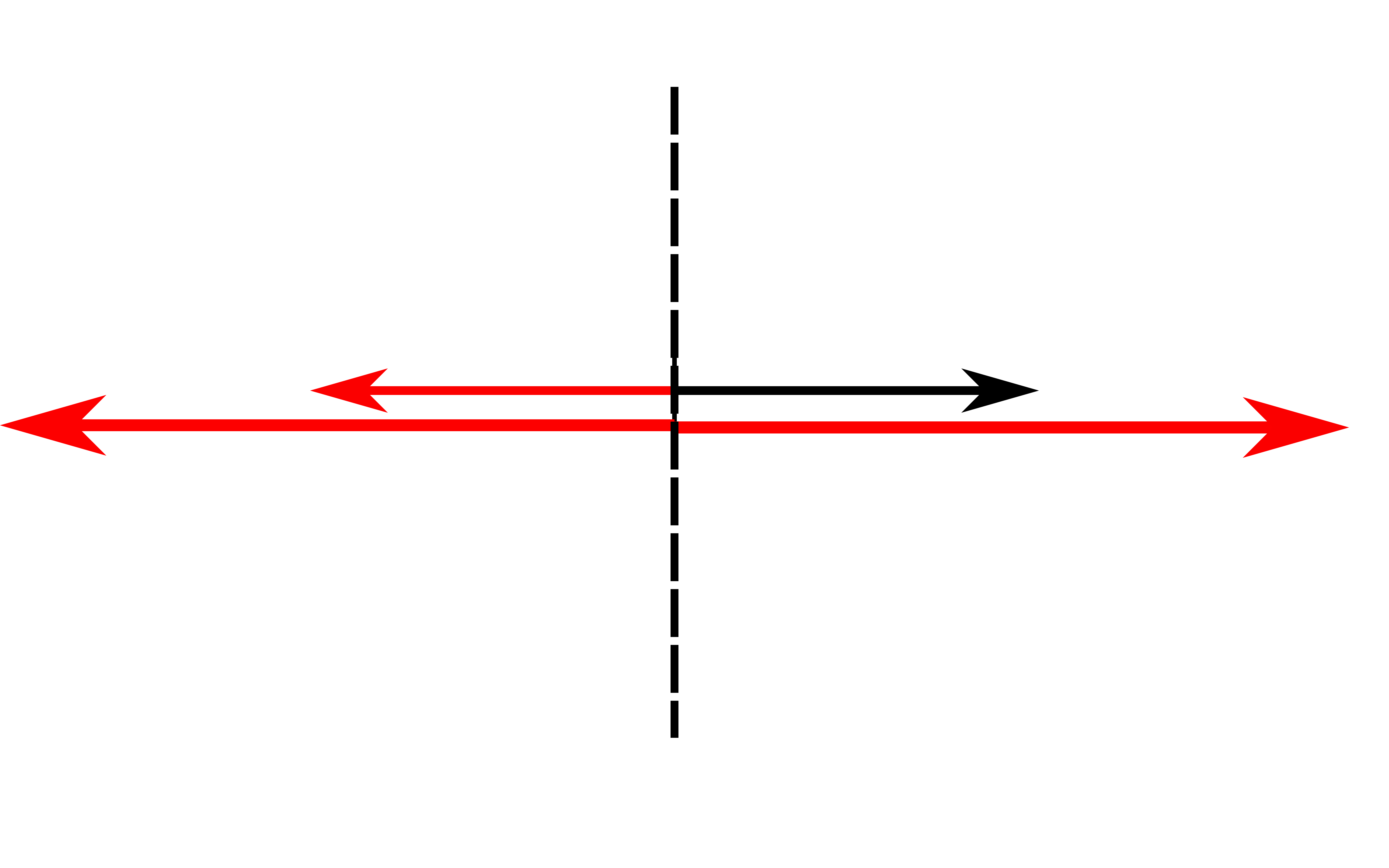}
\end{subfigure}\qquad\qquad
\begin{subfigure}{0.35\textwidth}
\def\svgwidth{\linewidth}{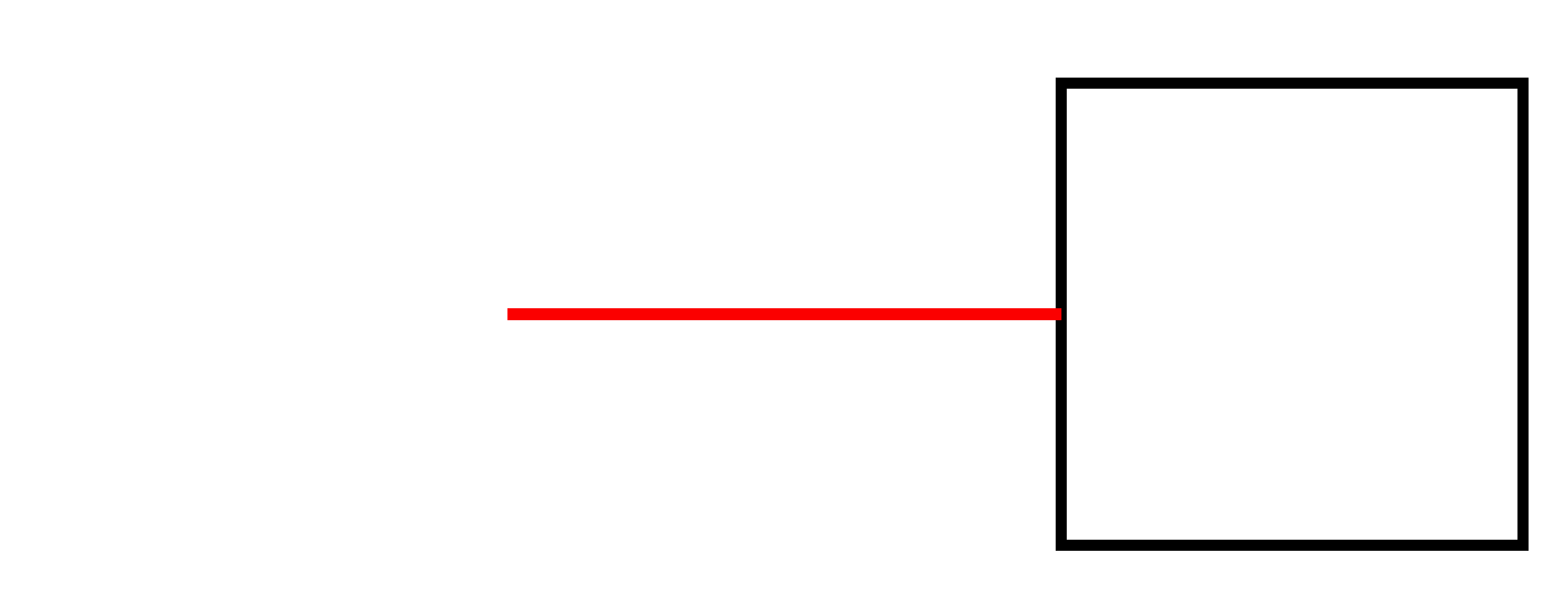}
\end{subfigure}
\caption{Configuration of charges and associahedron for the cyclic polytope $\mathcal{C}(4,2)$.}
\label{fig:associahedron24}
}
\end{figure}

In order to find the volume function $\Omega_{4,1}^{(2)}$, we follow the method outlined in Section \ref{Sec:mainstatement} and calculate the Jeffrey-Kirwan residue of the differential form
\begin{equation}
\omega_{4,1}^{(2)}(Y,Z) = \frac{1}{\langle 234 \rangle} \frac{ dx_1 }{ x_1 \left(\frac{\langle Y34 \rangle}{\langle 234 \rangle} - x_1 \frac{\langle 134 \rangle}{\langle 234 \rangle}\right)\left(- \frac{\langle Y24 \rangle}{\langle 234 \rangle} + x_1 \frac{\langle 124 \rangle}{\langle 234 \rangle}\right)\left(\frac{\langle Y23 \rangle}{\langle 234 \rangle} - x_1 \frac{\langle 123 \rangle}{\langle 234 \rangle}\right)}\,.
\end{equation}
From this form we can read off the set of one-dimensional charges
\begin{equation}
 B_{4,1}^{(2)}=\left\{\beta_1,\beta_2,\beta_3,\beta_4\right\}=\left\{ 1,-\frac{\langle 134 \rangle}{\langle 234 \rangle},\frac{\langle 124 \rangle}{\langle 234 \rangle},-\frac{\langle 123 \rangle}{\langle 234 \rangle}\right\}\,.
\end{equation}
Since we assumed that the external data is positive we find that $\beta_1,\beta_3>0$ and $\beta_2,\beta_4<0$, as shown in Fig.~\ref{fig:associahedron24}.
These charges define four one-dimensional cones:
\begin{equation}
\mathrm{Cone_{\{\beta_1\}}}=\mathrm{Cone_{\{\beta_3\}}}=\mathbb{R}_+\,,\qquad\quad \mathrm{Cone_{\{\beta_2\}}}=\mathrm{Cone_{\{\beta_4\}}}=\mathbb{R}_-\,.
\end{equation} 
The set of generic vectors with respect to $B_{4,1}^{(2)}$ is $\mathbb{R}\setminus\{0\}$ which has two connected components corresponding to the chambers: $\Lambda_1=\{\eta:\eta>0\} $ and $\Lambda_2=\{\eta:\eta<0\}$.
The computation of the Jeffrey-Kirwan residue for  the first chamber $\eta\in\Lambda_1$ gives\footnote{We will omit $B$ and $\eta$ in the Jeffrey-Kirwan residue notation when the charges and chamber vectors are clear from the context.}
\begin{eqnarray}
\mathrm{JKRes} \,\omega_{4,1}^{(2)} &=& \underset{\{\beta_1(x_1)=0\}}{\mathrm{JKRes}} \, \omega_{4,1}^{(2)} + \underset{\{\beta_3(x_1)=0\}}{\mathrm{JKRes}}\, \omega_{4,1}^{(2)}   = [234] + [241]\,,
\end{eqnarray}
while for the second chamber $\eta\in\Lambda_2$
\begin{eqnarray}
\mathrm{JKRes}\, \omega_{4,1}^{(2)} &=& \underset{\{\beta_2(x_1)=0\}}{\mathrm{JKRes}} \, \omega_{4,1}^{(2)} + \underset{\{\beta_4(x_1)=0\}}{\mathrm{JKRes}}\, \omega_{4,1}^{(2)} =[134] + [123]\,.
\end{eqnarray}
It is easy to notice using \eqref{GRTm2} that the results we got in both chambers agree.
Finally, we see that the secondary polytope $K_{3}$ is just an interval, presented in Fig.~\ref{fig:associahedron24}, where each vertex depicts one of the two only  possible ways of triangulating a square.

\bigskip
\noindent {\bf Example: $\mathbf{n=5}$.}

In this example, we start to see the full advantage of our method compared to the $i\epsilon$-prescription detailed in Example 7.29 of \cite{Arkani-Hamed:2017tmz}. In particular, in our case there is no need for studying the relative size of the $\epsilon$-parameters used there. 

\begin{figure}[ht]
\centering{
\begin{subfigure}{0.35\textwidth}
\def\svgwidth{\linewidth}{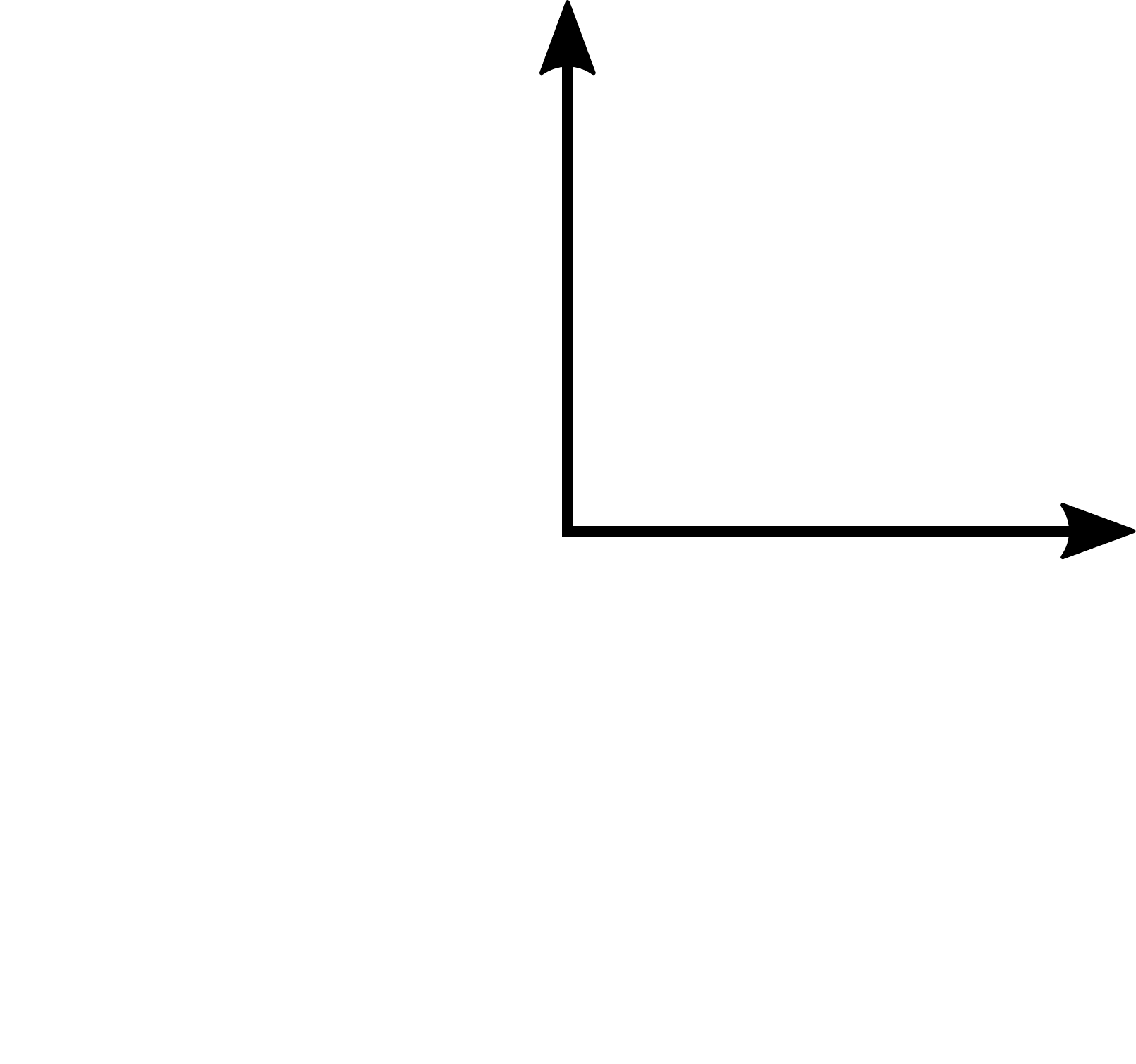}
\end{subfigure}\qquad\qquad
\begin{subfigure}{0.55\textwidth}
\def\svgwidth{\linewidth}{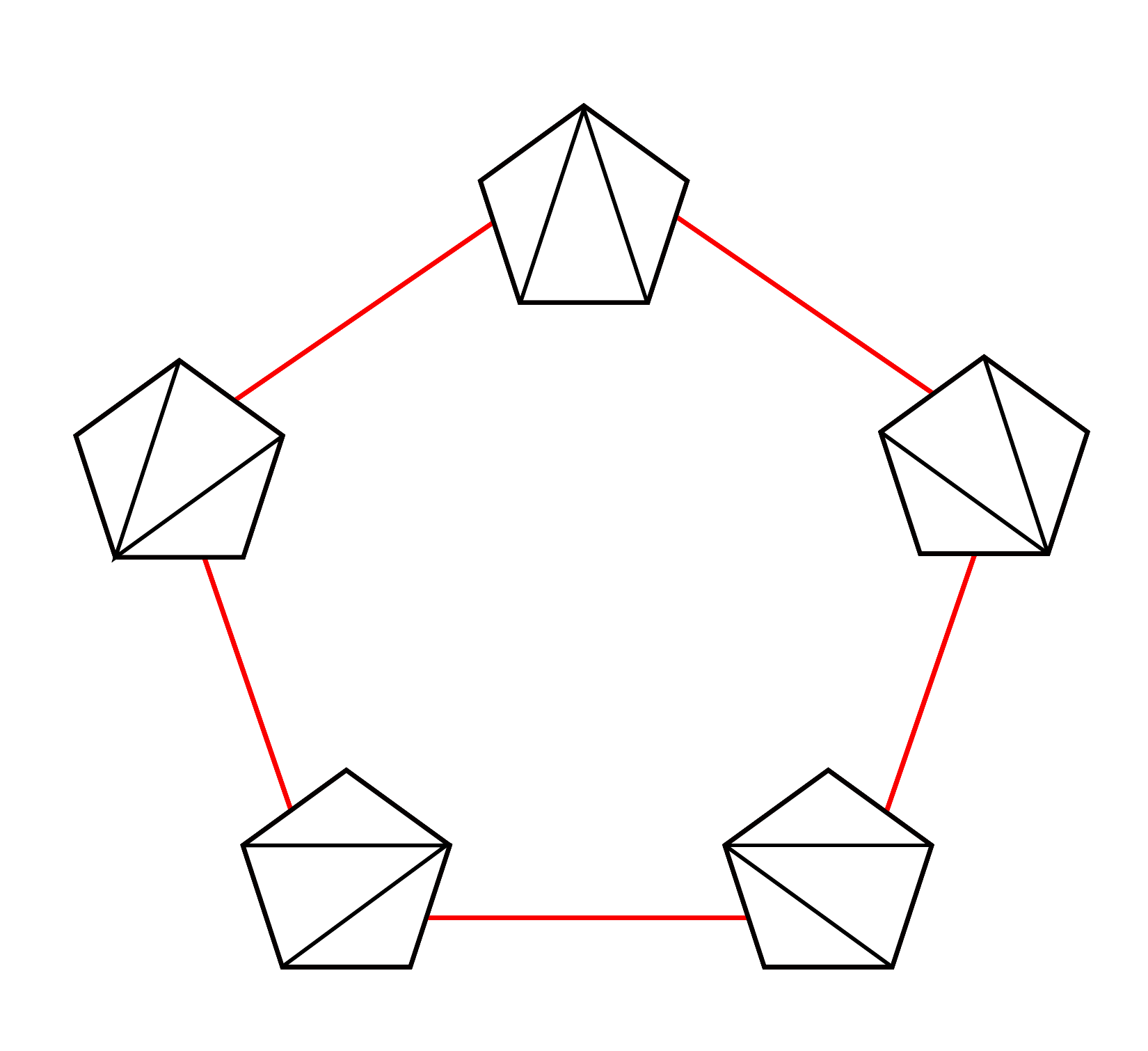}
\end{subfigure}
}
\caption{Configuration of charges and associahedron for the cyclic polytope $\mathcal{C}(5,2)$.}
\label{fig:associahedron25}
\end{figure}

The set of charges can be easily obtained from the  general formula \eqref{Zperp} and their explicit form is
\begin{equation}
B_{5,1}^{(2)}=\left\{(1,0),(0,1),\left(-\frac{\langle 145 \rangle}{\langle 345 \rangle}, -\frac{\langle 245 \rangle}{\langle 345 \rangle} \right) ,\left(\frac{\langle 135 \rangle}{\langle 345 \rangle},\frac{\langle 235 \rangle}{\langle 345 \rangle} \right),\left(-\frac{\langle 134 \rangle}{\langle 345 \rangle}, -\frac{\langle 234 \rangle}{\langle 345 \rangle} \right)\right\}\,.
\end{equation}
We depicted them in the two-dimensional space spanned by $\beta_1$ and $\beta_2$ in Fig.~\ref{fig:associahedron25}. Importantly, the positivity of the external data ensures that the vectors 
\begin{equation}\label{ordering521}
\{-\beta_1,\beta_2,-\beta_3,\beta_4,-\beta_5,\beta_1,-\beta_2,\beta_3,-\beta_4,\beta_5\}\,
\end{equation}
 are ordered clockwise. This implies that for any generic $\eta\in\mathbb{R}^2$ there are three cones containing $\eta$. 
We find that there are exactly five chambers, corresponding to the five connected components of the set of generic vectors. We label each chamber by the pair of charges which spans it, as depicted in Fig.~\ref{fig:associahedron25}. For each chamber we list all cones including it
\begin{align}\label{chambers52.1}
\Lambda_{14} &= \mathrm{Cone}_{\{\beta_1,\beta_2\}} \cap \mathrm{Cone}_{\{\beta_1,\beta_4\}} \cap \mathrm{Cone}_{\{\beta_3,\beta_4\}} \,,\\
\Lambda_{42} &= \mathrm{Cone}_{\{\beta_4,\beta_2\}} \cap \mathrm{Cone}_{\{\beta_1,\beta_2\}} \cap \mathrm{Cone}_{\{\beta_4,\beta_5\}} \,,\\
\Lambda_{25} &= \mathrm{Cone}_{\{\beta_2,\beta_5\}} \cap\mathrm{Cone}_{ \{\beta_4,\beta_5\}} \cap \mathrm{Cone}_{\{\beta_2,\beta_3\}} \,,\\
\Lambda_{53} &= \mathrm{Cone}_{\{\beta_5,\beta_3\}} \cap \mathrm{Cone}_{\{\beta_2,\beta_3\}} \cap \mathrm{Cone}_{\{\beta_5,\beta_1\}} \,,\\\label{chambers52.5}
\Lambda_{31} &= \mathrm{Cone}_{\{\beta_3,\beta_1\}} \cap \mathrm{Cone}_{\{\beta_5,\beta_1\}} \cap\mathrm{Cone}_{ \{\beta_3,\beta_4\}} \,.
\end{align}
Using the Jeffrey-Kirwan residue as in \eqref{omegafromJK}, we find five different representations of the volume functions:
\begin{align}\label{rep14}
\Lambda_{14}:\Omega_{5,1}^{(2)}&=[534]+[523]+[512]\,,\\\label{rep42}
\Lambda_{42}:\Omega_{5,1}^{(2)}&=[351]+[345]+[312]\,,\\
\Lambda_{25}:\Omega_{5,1}^{(2)}&=[134]+[123]+[145]\,,\\
\Lambda_{53}:\Omega_{5,1}^{(2)}&=[412]+[451]+[423]\,,\\\label{rep31}
\Lambda_{31}:\Omega_{5,1}^{(2)}&=[245]+[234]+[251]\,.
\end{align}
All these expressions represent the same function and they are related to each other by global residue theorems. For example, if we take neighbouring chambers, e.g.~$\Lambda_{14}$ and $\Lambda_{42}$, then the two representations \eqref{rep14} and \eqref{rep42} are related to each other by the bistellar flip
\begin{equation}
[523]+[512]=[351]+[312]\,.
\end{equation}

\begin{figure}[ht]
\begin{center}\includegraphics[scale=0.51]{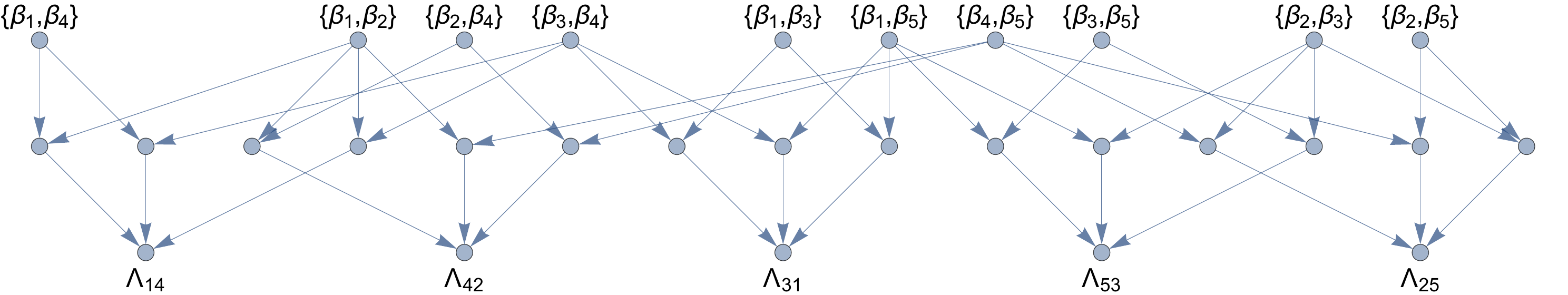}\end{center}\caption{Hasse diagram for the partial order of cone intersection in the secondary fan of the cyclic polytope $\mathcal{C}(5,2)$.}
\label{fig:poset521}
\end{figure}

As final part of this example, we discuss the Hasse diagram of the poset $(\Lambda,\prec)$ defined in \eqref{partial.order}, depicted in Fig.~\ref{fig:poset521}. Vertices in the first row of the picture correspond to all possible cones, each labelled by a pair of $\beta$'s. The second and third row indicate all non-trivial double and triple intersections of cones, respectively. One can easily find all chambers we discussed in \eqref{chambers52.1}-\eqref{chambers52.5} by looking at all vertices with no outgoing arrow. The poset $(\Lambda,\prec)$ provides an extremely useful way to find all chambers in more complicated examples when the secondary fan is difficult to visualize in high-dimensional spaces.

\bigskip
\noindent {\bf Higher $\mathbf{n}$.}

When $n>5$, there are at least two new features appearing for $m=2$. Firstly, one notices that for $n=4,5$ the chambers always coincide with particular cones. This comes from the fact that in one and two dimensions all cones are simplicial. This is not true any more for higher number of points where the space of charges is higher-dimensional and chambers start to be polyhedral. In that case the algorithmic procedure for finding chambers becomes crucial if one is interested in finding the complete list of triangulations. Secondly, one notices that, up to a cyclic relabelling, all triangulations for $n=4,5$ are of the form $\sum\limits_i[1 \,i\, i+1]$. Starting from $n=6$ we also encounter triangulations which are of different form, for example
\begin{equation}
\Omega_{6,1}^{(2)}=[123]+[135]+[156]+[345]\,.
\end{equation} 


\subsection{Physical Amplituhedron: \texorpdfstring{$\mathbf{m=4}$}{}.}
We continue with the study of the physical tree amplituhedron, which is relevant for scattering amplitudes in planar $\mathcal{N}=4$ SYM. The external data defines a convex polytope in $\mathbb{P}^4$. 
Similarly to the $m=2$ case, we can derive general relations generating all  global residue theorems, namely the following six-term identity 
\begin{equation}\label{GRTm4}
[i_1i_2i_3i_4i_5]+[i_1i_2i_3i_5i_6]+[i_1i_3i_4i_5i_6]=[i_1i_2i_3i_4i_6]+[i_1i_2i_4i_5i_6]+[i_2i_3i_4i_5i_6]\,,
\end{equation}
for any set of indices $1\leq i_1<i_2<i_3<i_4<i_5<i_6\leq n$. Geometrically, these relations define bistellar flips in four dimensions, allowing us to relate triangulations in adjacent chambers. The triangulations related by a bistellar flip \eqref{GRTm4} have the same number of simplices and therefore each chamber in the JK-residue prescription is contained in the same number of cones, namely $(n-3)(n-4)/2$.

\bigskip
\noindent {\bf Example: $\mathbf{n=6}$.}

We start by writing the explicit form of charges
\begin{equation}
 B_{6,1}^{(4)}=\left\{ 1,-\frac{\langle 13456 \rangle}{\langle 23456 \rangle},\frac{\langle 12456 \rangle}{\langle 23456 \rangle},-\frac{\langle 12356 \rangle}{\langle 23456 \rangle},\frac{\langle 12346 \rangle}{\langle 23456 \rangle},-\frac{\langle 12345 \rangle}{\langle 23456 \rangle}\right\}\,.
\end{equation}
Importantly, from positivity of external data, we find that $\beta_1,\beta_3,\beta_5$ are positive and $\beta_2,\beta_4,\beta_6$ are negative. The space of charges is divided in two chambers: $\Lambda_1=\mathbb{R}_+$ and $\Lambda_2=\mathbb{R}_-$. The Jeffrey-Kirwan residue prescription for $\eta\in\Lambda_1$ gives
\begin{eqnarray}
\mathrm{JKRes} \,\omega_{6,1}^{(4)}  = [23456] + [12456]+[12346]\,,
\end{eqnarray}
while for $\eta\in\Lambda_2$
\begin{eqnarray}
\mathrm{JKRes} \,\omega_{6,1}^{(4)}  = [13456] + [12356]+[12345]\,.
\end{eqnarray}
It is straightforward to check that these two answers are equal by applying the bistellar flip \eqref{GRTm4}. Similar to the case $n=4,m=2$, the secondary polytope is a segment.

\bigskip
\noindent {\bf Higher $\mathbf{n}$.}

For $n=7$ there are seven chambers corresponding to seven distinguished triangulations and the secondary polytope is a heptagon in two dimensions. All triangulations are of the form $\sum\limits_{i<j}[1\,i\, i+1\, j\, j+1]$ and its cyclic permutations. For $n=8$ there are 40 triangulations, each with exactly ten triangles, and the three-dimensional secondary polytope is depicted e.g.~in Fig.~13 of \cite{PfeifleRambau}. It implies that there are exactly 40 different ways to write the volume function $\Omega_{8,1}^{(4)}$ as a sum of ten brackets \eqref{brackets}, all these representations easily obtained from our JK-residue prescription. For higher number of points, the number of triangulations grows rapidly and it can be found up to $n=12$ in Table 1 of \cite{RAMBAU200065}.  

\subsection{Odd-Dimensional Cyclic Polytopes}
\label{sec:odd.cyclic}
\begin{figure}[ht!]
\centering{\def\svgwidth{0.7\linewidth}{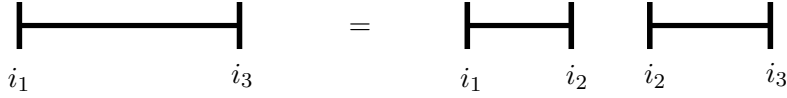}
\caption{Global residue theorem or bistellar flip for $m=1$.}
\label{fig:bistellarm1}
}
\end{figure}

There is a significant difference for odd-dimensional cyclic polytopes compared to our discussion for even dimensions. We have observed that, for the even-dimensional case, all triangulations have the same number of simplices. This descends from the fact that bistellar flips involve even number of terms which are divided in two groups with the same number of elements. This does not apply in odd dimensions. In particular, the bistellar flip for $m=1$ is depicted in Fig.~\ref{fig:bistellarm1} and it takes the form
\begin{equation}\label{bistellar1}
[i_1 i_3]=[i_1 i_2]+[i_2 i_3]\,.
\end{equation}

\begin{figure}[ht]
\centering{
\def\svgwidth{0.7\linewidth}{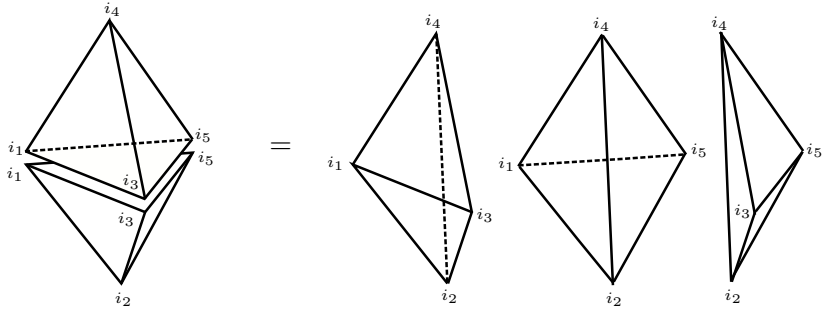}
\caption{Global residue theorem or bistellar flip for $m=3$.}
\label{fig:bistellarm3}
}
\end{figure}

\noindent For $m=3$ the bistellar flip is
\begin{equation}
[i_1 i_2i_3i_5]+[i_1i_3i_4i_5]=[i_1i_2i_3i_4]+[i_1i_2i_4i_5]+[i_2i_3i_4i_5]\,,
\end{equation}
which geometrically can be depicted as in Fig.~\ref{fig:bistellarm3}.
Therefore, bistellar flips do not preserve the number of triangles in triangulations. From the point of view of the JK-residue this is manifested by the fact that chambers can have a different number of cones containing them.

\bigskip
\noindent{{\bf Example: $\mathbf{m=1}$}

For $m=1$ the positive external data define an ordered set of points in $\mathbb{P}^1$. The secondary polytope has been studied in \cite{gelfand2008discriminants} and all triangulations are given by subdivisions of the interval $[1,n]$ into smaller intervals of the form $[i,j]$, $i<j$. Therefore, the triangulations are labelled by subsets of $\{2,3,\ldots,n-1\}$. We show how this description arises from the chamber structure in the space of charges. For $n=3$ the charges are
\begin{equation}
B_{3,1}^{(1)}=\left\{1,-\frac{\langle 13\rangle}{\langle 23\rangle},\frac{\langle 12\rangle}{\langle 23\rangle}\right\}\,,
\end{equation}
and we have two chambers. The secondary polytope is a segment and the two triangulations are: $\{[1,3]\}$ and $\{[1,2],[2,3]\}$, related by the bistellar flip \eqref{bistellar1}. In the case $n=4$ the charges
\begin{equation}
B_{4,1}^{(1)}=\left\{(1,0),(0,1),\left(-\frac{\langle 14\rangle}{\langle 34\rangle},-\frac{\langle 24\rangle}{\langle 34\rangle}\right),\left(\frac{\langle 13\rangle}{\langle 34\rangle},\frac{\langle 23\rangle}{\langle 34\rangle}\right)\right\}
\end{equation}
are depicted in Fig.~\ref{fig:associahedron14}. There are four chambers and the secondary polytope is a quadrilateral. For $n=5$ the list of charges can be read off from \eqref{Zperp}: the dual space is divided in eight chambers and the secondary polytope depicted in Fig.~\ref{fig:associahedron15} is combinatorially equivalent to a cube. In general, for any $n$ there are $2^{n-2}$ chambers and the secondary polytope is combinatorially equivalent to a hypercube in $n-2$ dimensions.
\begin{figure}[ht]
\centering{
\begin{subfigure}{0.3\textwidth}
\def\svgwidth{\linewidth}{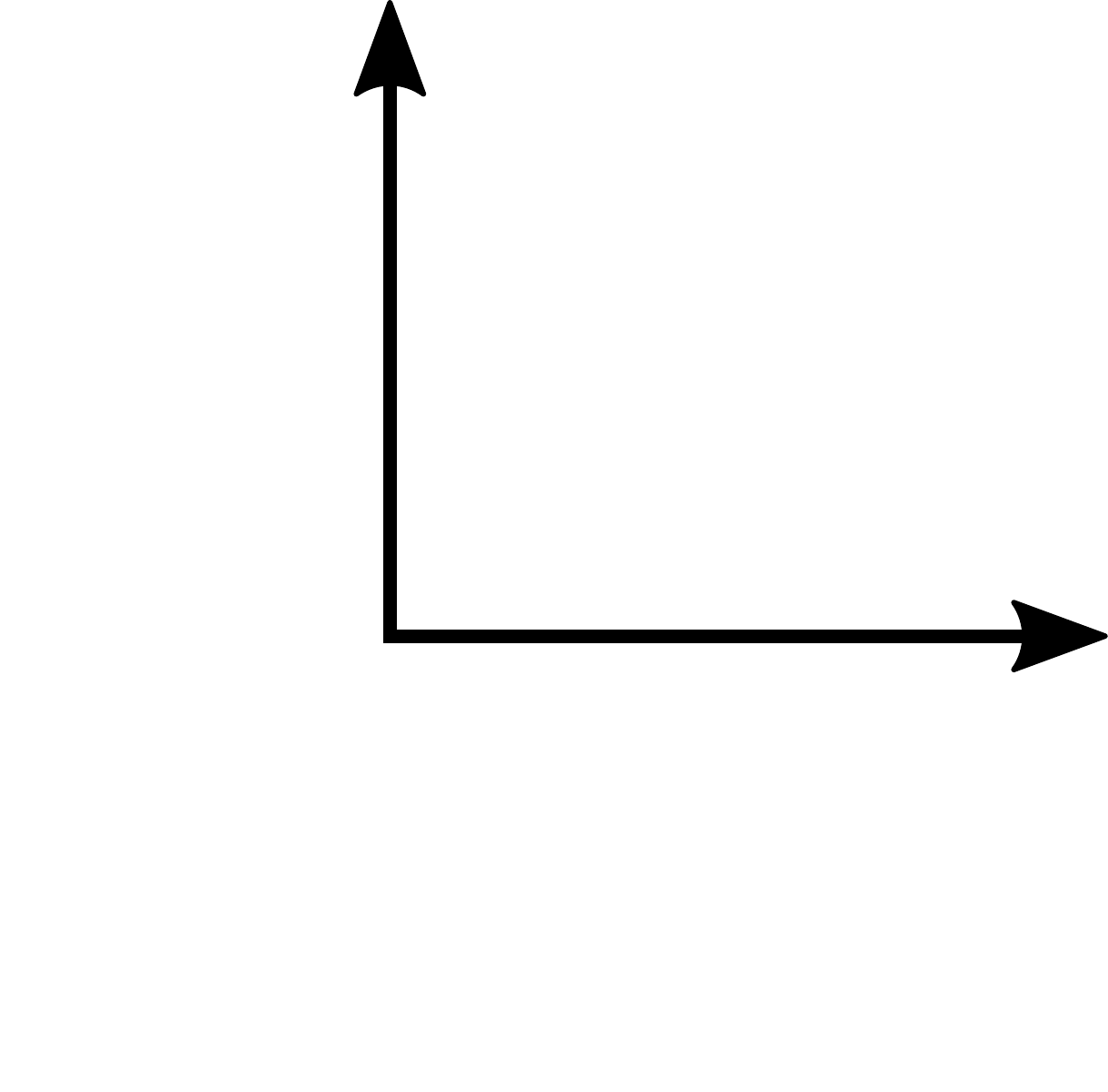}
\end{subfigure}\qquad\qquad
\begin{subfigure}{0.4\textwidth}
\def\svgwidth{\linewidth}{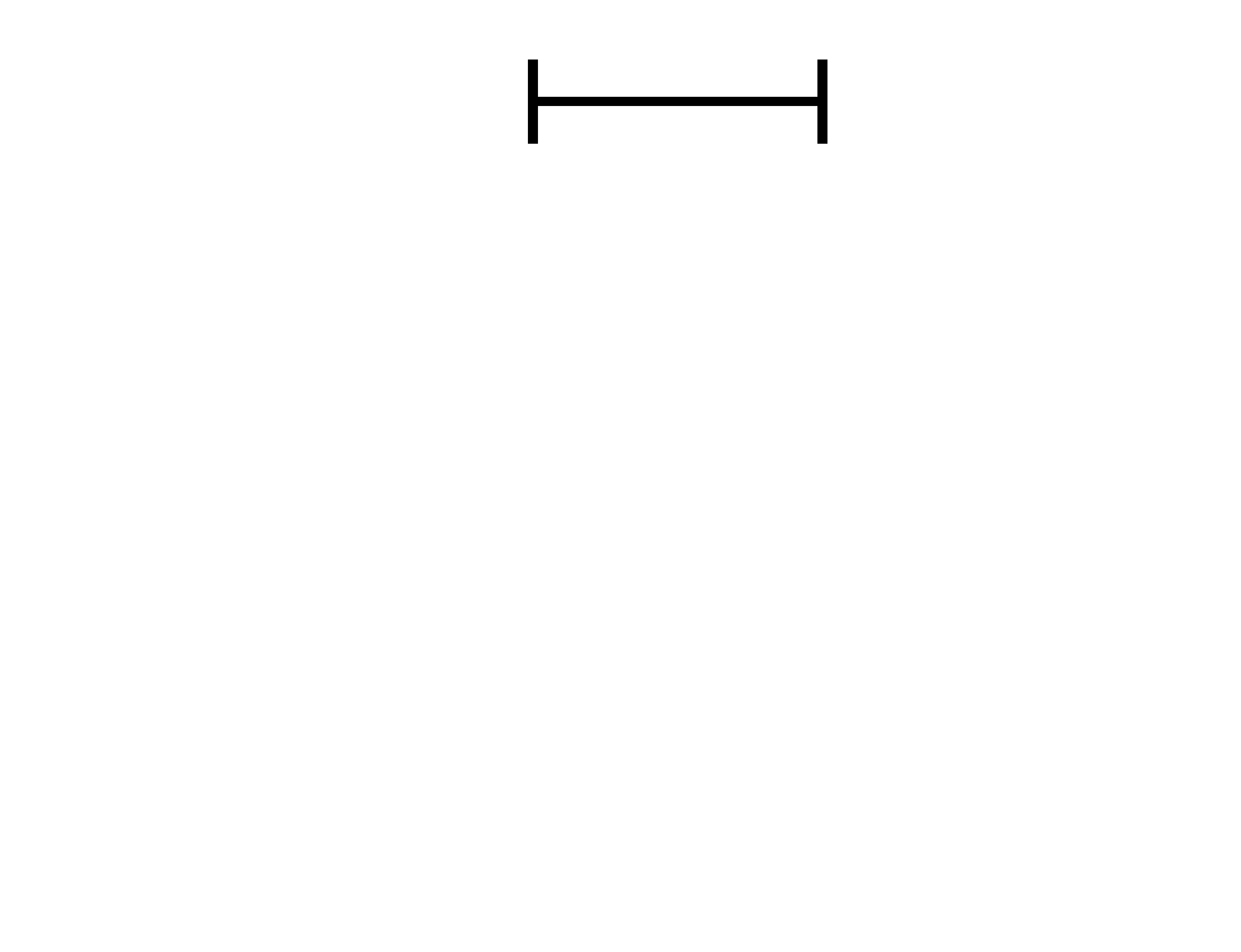}
\end{subfigure}
\caption{Configuration of charges and secondary polytope for the cyclic polytope $\mathcal{C}(4,1)$.}
\label{fig:associahedron14}
}
\end{figure}
\begin{figure}[ht]
\centering{
\def\svgwidth{0.6\linewidth}{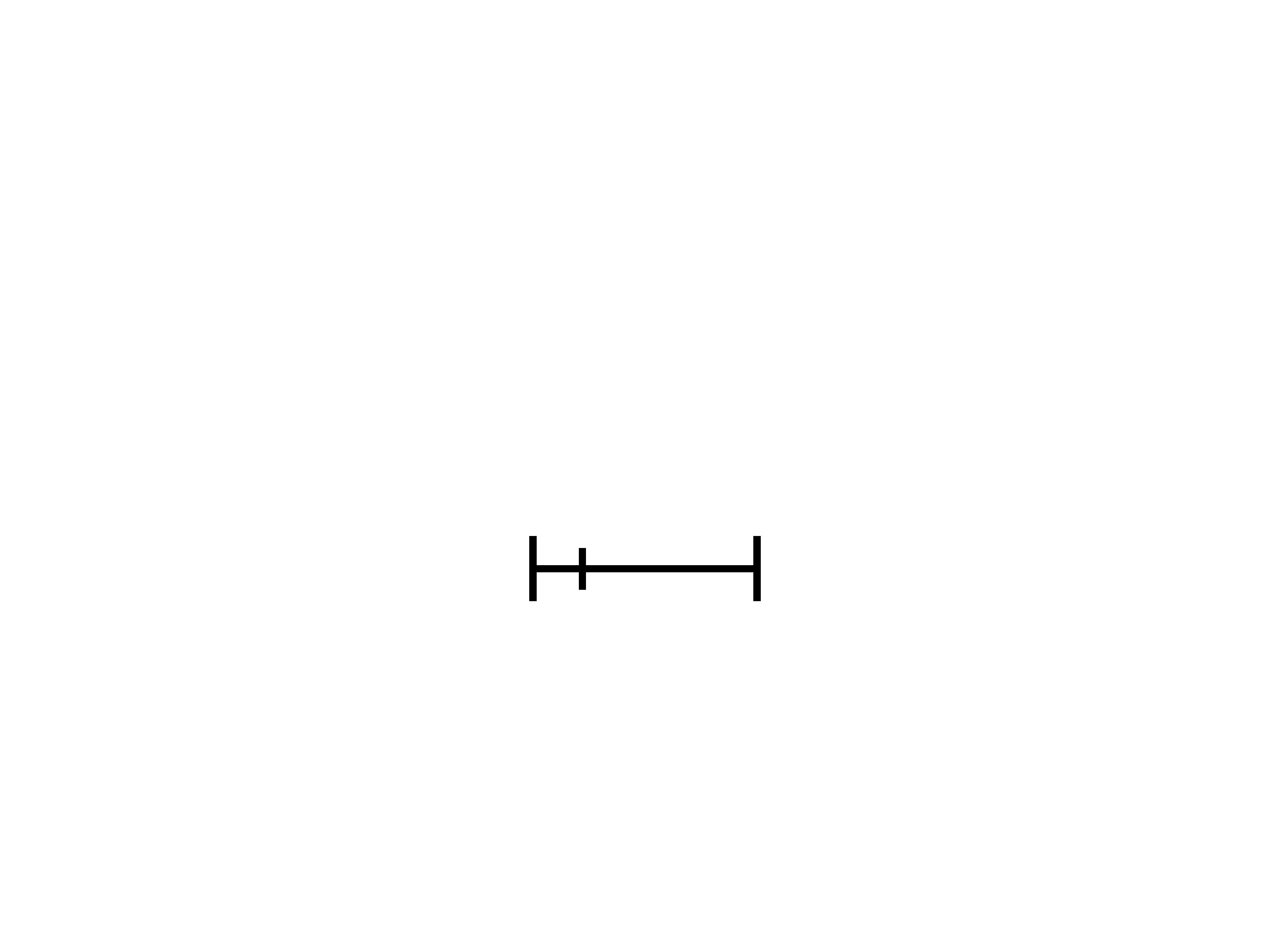}
\caption{Secondary polytope for the cyclic polytope $\mathcal{C}(5,1)$.}
\label{fig:associahedron15}
}
\end{figure}

\bigskip
\noindent {\bf Example: $\mathbf{m=3}$}

By now we observe a pattern of triangulations for cyclic polytopes with small number of external points: for $n=m+2$ the secondary polytope is always a segment and for $n=m+3$ it is an $n$-gon. For $m=3$ this implies that for $n=5$ the secondary polytope is a segment and two triangulations are given by either side of the bistellar flip depicted in Fig.~\ref{fig:bistellarm3}. For $n=6$ the secondary polytope is a hexagon. For general $n=m+4$ secondary polytopes are three-dimensional and start to be quite complicated. For $m=3$, $n=7$ we find that there are exactly 25 triangulations and the secondary polytope is depicted in Fig.~\ref{fig:secondary37}. As we have pointed out already, the number of simplices in each triangulation differs for odd $m$ and in the figure we have indicated how many simplices are in each triangulation. Notice that each edge connects triangulations for which the number of simplices differs by exactly one. This descends from the fact that each such pair of triangulations is related to each other by a bistellar flip. 
 \begin{figure}[ht]
\centering{\includegraphics[scale=0.3]{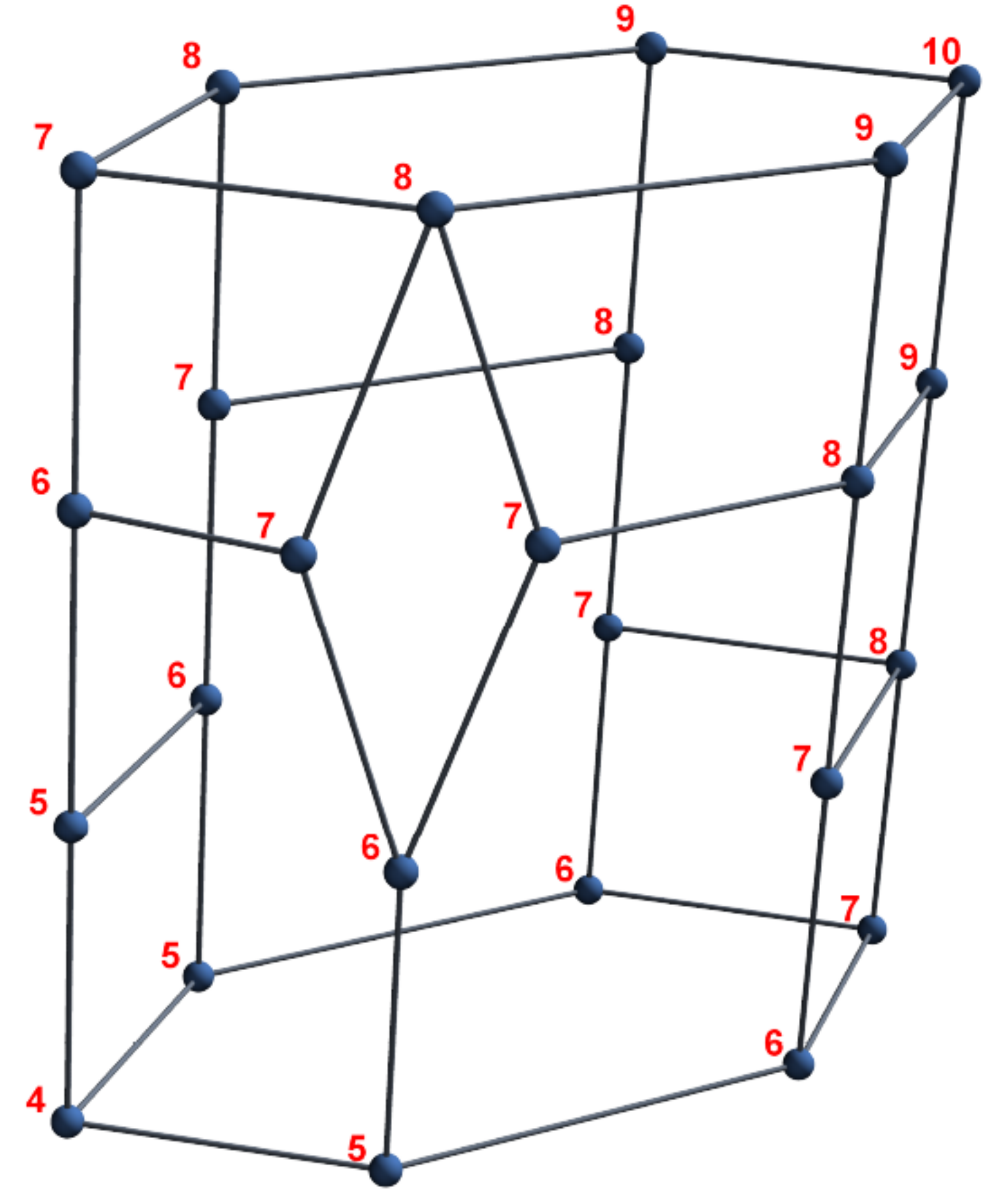}
\caption{Secondary polytope for the cyclic polytope $\mathcal{C}(7,3)$. The labels indicate how many simplices there are in a given triangulation.}
\label{fig:secondary37}
}
\end{figure}

\subsection{Conjugate Cyclic Polytopes: \texorpdfstring{$\mathbf{n-m-k=1}$}{}}

In this section we describe how the JK-residue is used to find volume functions for amplituhedra conjugate to cyclic polytopes. As before, we will provide various possible representations of a given volume function, which are related to distinguished triangulations of the amplituhedron $\mathcal{A}_{n,n-m-1}^{(m)}$. We observe that the structure of triangulations for even $m$ is identical for cyclic polytopes and their conjugates. On the other hand, for odd $m$, we argue that the conjugate amplituhedron is combinatorially different from the corresponding cyclic polytope and, in particular, we find that the JK-residue procedure applied to the integral \eqref{omegaintegralconjugate} does not give the correct volume form. This also contradicts the statement in \cite{Arkani-Hamed:2017tmz}, where the $i\epsilon$-prescription for conjugate amplituhedra was claimed to work also for odd $m$. 

We start by taking $m=2$ and focusing first on the simplest example with $n=5$, which implies that $k=2$. This is the conjugate case to the $n=5,m=2,k=1$ example presented in Section \ref{Sec:toy}. As we explained in Sec.~\ref{Sec:mainstatement}, in order to find charges one solves the $\delta$-functions in \eqref{omegaintegralconjugate} and change variables from the $c$-variables to the minors. We keep $c_{\underline{1}}=(12)$ and $c_{\underline{2}}=(23)$ as the unknowns which leads to the following set of charges:
\begin{align}
B_{5,2}^{(2)}&=\left\{\beta_{(12)},\beta_{(23)},\beta_{(34)},\beta_{(45)},\beta_{(15)}\right\} \\\nonumber
&=\left\{(1,0),(0,1),\left( -\frac{\langle Y 15\rangle\langle 1245\rangle}{\langle Y 45\rangle\langle 1345\rangle},\frac{\langle Y (125)\cap(345)\rangle}{\langle Y 45\rangle\langle 1345\rangle}\right),\right.\\ &\left.
\left( -\frac{\langle Y (123)\cap(145)}{\langle Y 45\rangle\langle 1345\rangle},\frac{\langle Y (123)\cap(345)\rangle}{\langle Y 45\rangle\langle 1345\rangle}\right),
\left( \frac{\langle Y (234)\cap(145)}{\langle Y 45\rangle\langle 1345\rangle},-\frac{\langle Y 34\rangle\langle 2 345\rangle}{\langle Y 45\rangle\langle 1345\rangle}\right)
 \right\}\,,
 \end{align}
 where $\langle Y(i_1i_2i_3)\cap(j_1j_2j_3)\rangle=\langle Y_1 i_1i_2i_3\rangle \langle Y_2 j_1j_2j_3\rangle-\langle Y_1j_1j_2j_3\rangle\langle Y_2 i_1i_2i_3\rangle$.
Notice that the last entry is labelled as $\beta_{(15)}$ which descends from the fact that we have $c_{\underline{5}}=(15)$.
\begin{figure}[ht]
\centering{
\def\svgwidth{0.4\linewidth}{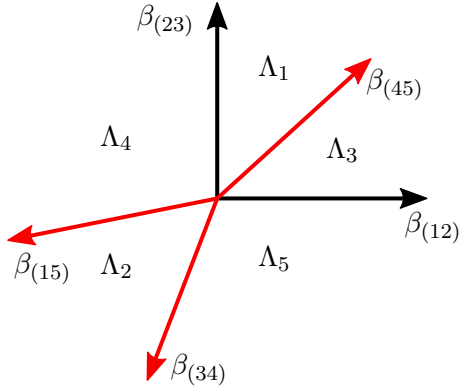}
\caption{Configuration of charges for  $n=5, m=2, k=2$.}
\label{fig:lattice25k2}
}
\end{figure}

Using positivity of data, we find relations between the charges similar to \eqref{ordering521}, namely the vectors
\begin{equation}
\{-\beta_{(12)},\beta_{(23)},-\beta_{(34)},\beta_{(45)},-\beta_{(15)},\beta_{(12)},-\beta_{(23)},\beta_{(34)},-\beta_{(45)},\beta_{(15)}\}
\end{equation}
are ordered clockwise. This leads to the configuration of charges in Fig.~\ref{fig:lattice25k2}, and therefore the same distributions of cones and chambers as for the case $n=5,m=2,k=1$. Different representations of the volume function $\Omega_{5,2}^{(2)}$ can then be found by applying formula \eqref{omegaconjugatefromJK} for generic, with respect to $B_{5,2}^{(2)}$, vectors. By varying over all chambers in  Fig.~\ref{fig:lattice25k2}, we obtain that $\Omega_{5,2}^{(2)}$ can be written as in formulas \eqref{rep14}-\eqref{rep31} with all brackets replaced by their conjugate counterparts \eqref{conjugatebracket}. The secondary polytope is again a pentagon which we depicted in Fig.~\ref{fig:associahedron25k2}. 
\begin{figure}[ht]
\centering{
\def\svgwidth{0.75\linewidth}{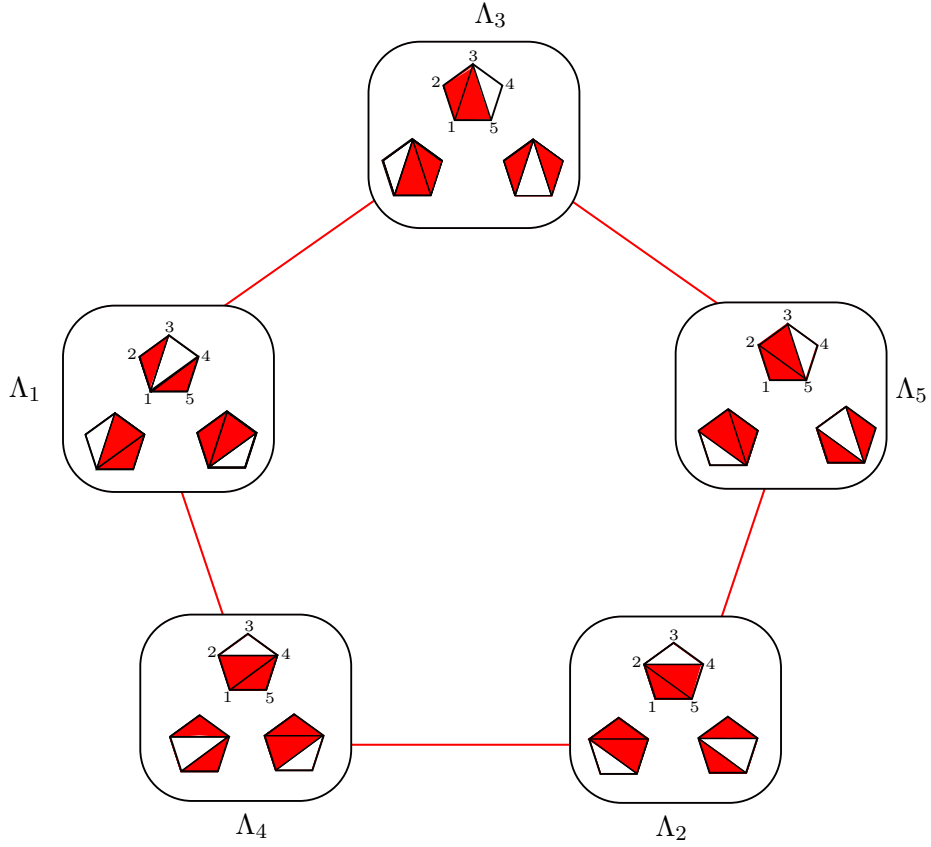}
\caption{Secondary polytope for $n=5, m=2, k=2$.}
\label{fig:associahedron25k2}
}
\end{figure}
Each vertex of a pentagon corresponds to a triangulation of $\mathcal{A}_{5,2}^{(2)}$ which consists of exactly three generalized triangles. For $m=2$ there exists a simple parametrization of each generalized triangle that allows for their graphical representation conjugate to the $k=1$. As an example let us consider $k=2$: each generalized triangle is the image of a four-dimensional Grassmannian cell parametrized by matrices for which the only non-vanishing entries are:
\begin{equation}
C_{\{\{i_1,i_2,i_3\},\{j_1,j_2,j_3\}\}}:\begin{cases}(c_{1 i_1},c_{1 i_2},c_{1 i_3})=(1,\star,\star)\,,\\
(c_{2 j_1},c_{2 j_2},c_{2 j_3})=(1,\star,\star)\,,\end{cases}
\end{equation}
and the sets $\{i_1,i_2,i_3\}$ and $\{j_1,j_2,j_3\}$ are vertices of non-intersecting triangles. This allows us to depict them as a collection of two non-intersecting triangles inside an $n$-gon. The construction easily generalizes to any $k$. Using this geometric representation we can depict each triangulation of the amplituhedron $\mathcal{A}_{5,2}^{(2)}$ as in Fig.~\ref{fig:associahedron25k2}. Similarly as in the $k=1$ case, the triangulations in neighbouring chambers can be related to each other by  a global residue theorem which for $m=2$ is a four-term identity conjugate to \eqref{GRTm2}
\begin{equation}
\overline{[i_{{1}}\, i_{{2}}\,i_{{3}}]}+\overline{[i_{{1}} \, i_{{3}}\, i_{{4}}]}=\overline{[i_{{2}}\,i_{{3}}\,i_{{4}}]}+\overline{[i_{{1}}\, i_{{2}}\,i_{{4}}]}\,, \qquad 1\leq i_1<i_2<i_3<i_4\leq n\,.
\end{equation}
The identity can be depicted as in Fig.~\ref{fig:GRTm2bar}, where the coloured circle indicates that the four-gon labelled by $i_1,\ldots,i_4$ is embedded in an $n$-gon.
\begin{figure}[ht]
\centering{
\def\svgwidth{0.8\linewidth}{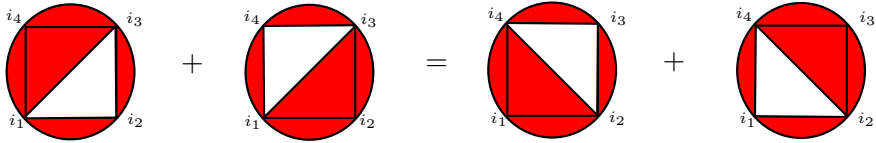}
\caption{Global residue theorem for $ m=2$ and $n-m-k=1$.}
\label{fig:GRTm2bar}
}
\end{figure}

For $m=2$ and general $n=k+3>5$ the secondary polytope is combinatorially equivalent to the associahedron $K_{n-1}$. Each triangulation of $\mathcal{A}_{n,n-3}^{(2)}$ can be easily obtained from the $k=1$ triangulations by replacing
\begin{equation}
[i_1\,i_2\,i_{3}] \to \overline{[{i}_{1}\,{i}_{2}\, {i}_{3}]}\,.
\end{equation}
Similarly, one can show that for $m=4$ the JK-residue prescription for conjugate amplituhedron leads to the same cone and chamber configuration as for the associated cyclic polytope. Also the global residue theorem is just the conjugation of \eqref{GRTm4}. Therefore, the secondary polytope of $\mathcal{A}_{n,1}^{(4)}$ can again be mapped to the secondary polytope of $\mathcal{A}_{n,n-5}^{(4)}$. We have checked this statement extensively, for various $n$.

We end this section by commenting on the conjugation for odd-dimensional amplituhedra. In this case the JK-residue prescription presented above for even $m$ does not seem to result in the correct volume function for odd $m$. The simplest example we can consider to examine this problem is $n=4,m=1$: in Section \ref{sec:odd.cyclic} we have already studied the cyclic polytope $\mathcal{C}(4,1)$. In particular, we found its four triangulations and all generalized triangles, which are the following six segments: $[1,2],[1,3],[1,4],[2,3],[2,4],[3,4]$. An interesting property of $\mathcal{C}(4,1)$ is that there exists a single segment, namely $[1,4]$, which covers the amplituhedron completely, i.e.~there is a triangulation with just one element. 
Let us now consider its conjugate $\mathcal{A}_{4,2}^{(1)}$. This amplituhedron was studied in details in \cite{Karp:2016uax}, where in particular it was shown that $\mathcal{A}_{n,k}^{(1)}$ can be identified with the complex of bounded faces of a cyclic hyperplane arrangement. We depict the case $n=4, k=2$ in Fig.~\ref{fig:n4m1k2}. 
\begin{figure}[htp]
    \centering
    \begin{subfigure}[b]{0.23\linewidth}
      \centering
 \def\svgwidth{1.1\linewidth}{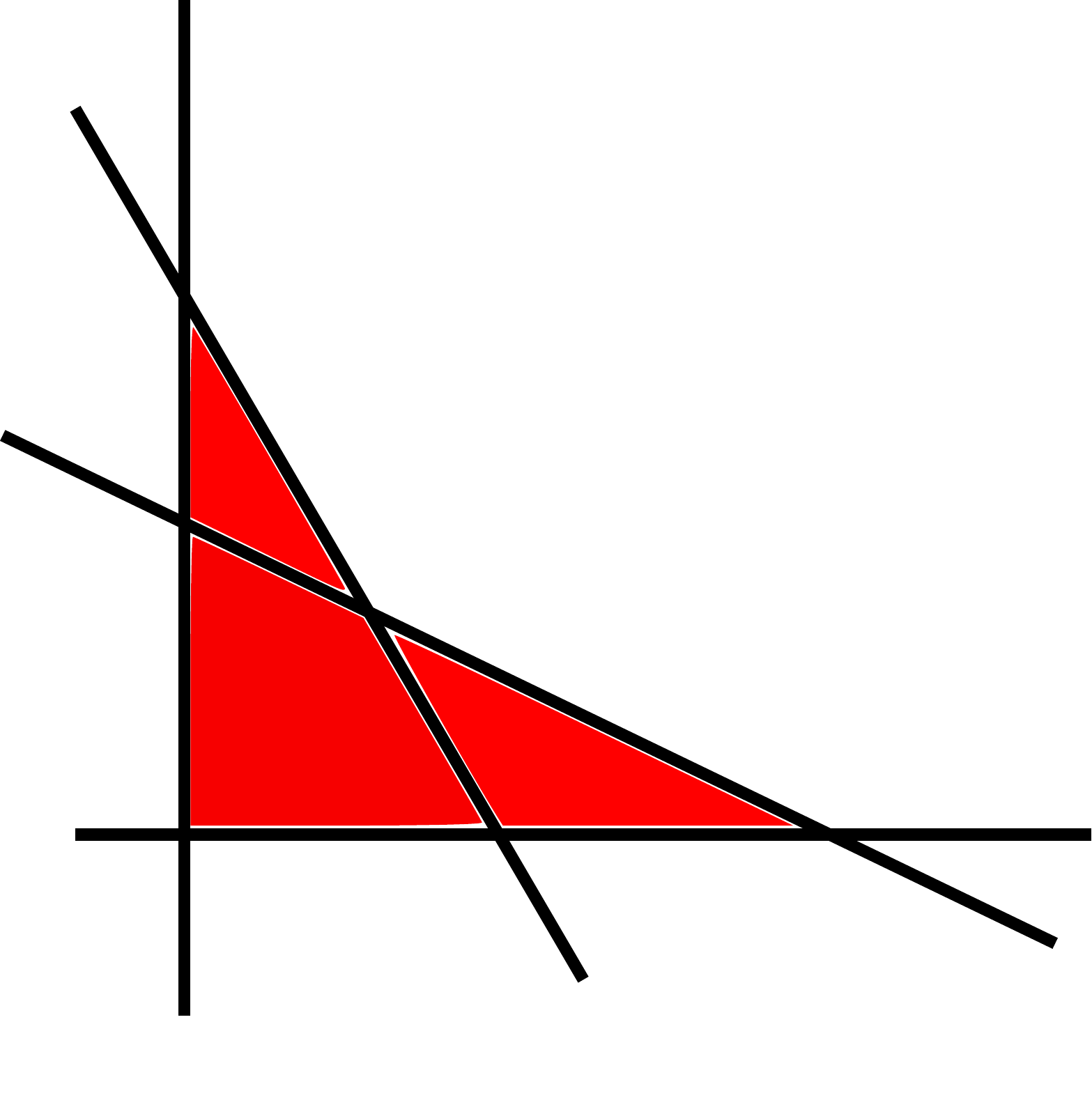}
      \caption{}
      \label{fig:n4m1k2}
    \end{subfigure}\qquad\quad
    \begin{subfigure}[b]{0.53\textwidth}
      \centering
      \qquad\qquad
     \includegraphics[scale=0.14]{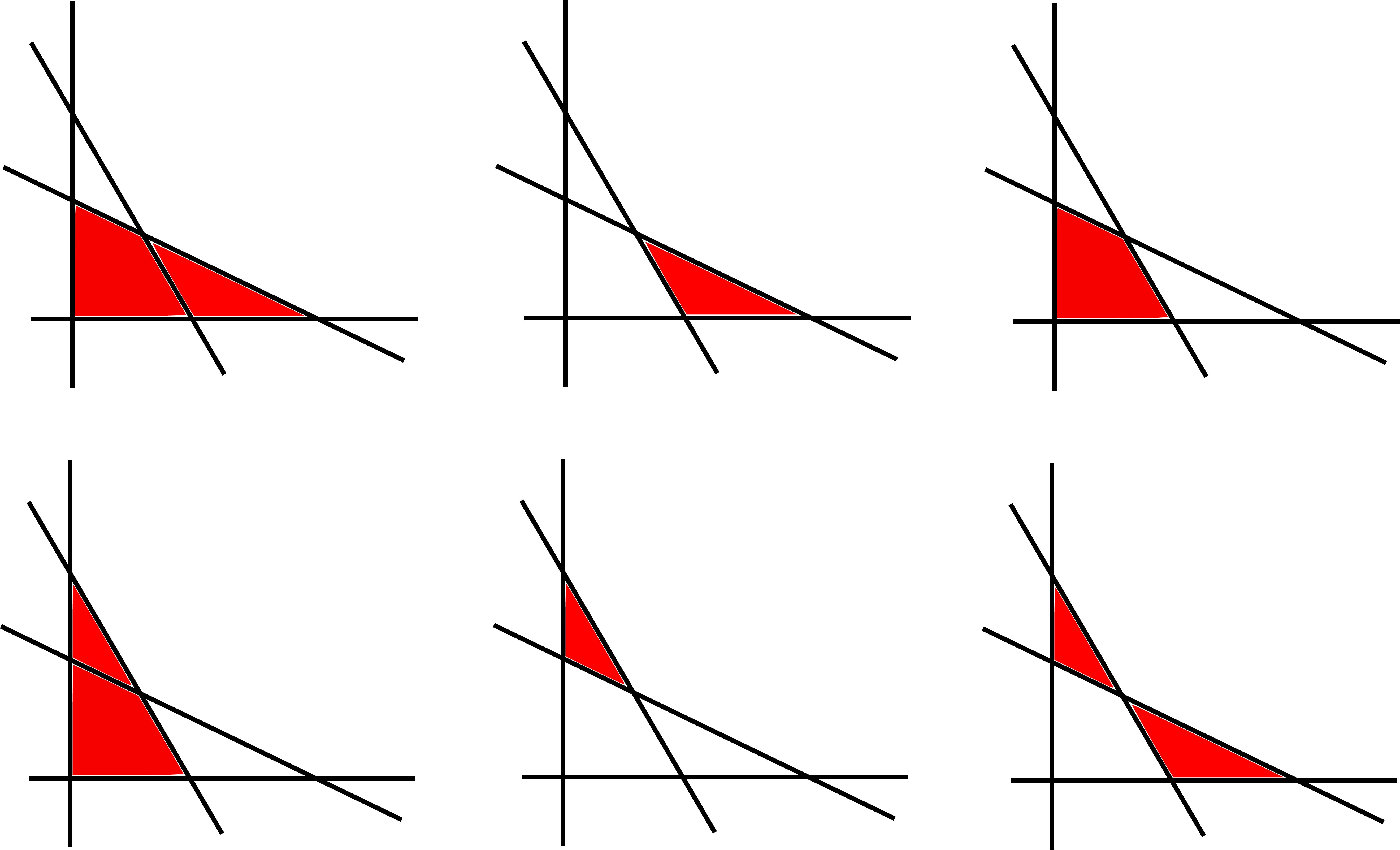}
      \caption{}
      \label{fig:gentrian}
    \end{subfigure}
  \caption {(a) The $n=4$, $m=1$, $k=2$ amplituhedron; (b) Generalized triangles for $\mathcal{A}_{4,2}^{(1)}$.}
  
\end{figure}
There are exactly six two-dimensional cells of the positive Grassmannian $G_+(4,2)$ which have top-dimensional images in the amplituhedron space, see Fig.~\ref{fig:gentrian}: four triangles bounded by the lines $\{\ell_1,\ell_2,\ell_3\}$, $\{\ell_1,\ell_2,\ell_4\}$, $\{\ell_1,\ell_3,\ell_4\}$ and $\{\ell_2,\ell_3,\ell_4\}$, respectively, the four-gon and the union of two triangles.  Importantly, it is clear that no single image covers the full amplituhedron. On the other hand, when we follow our JK-residue prescription we find a configuration of charges which looks similar to the one in Fig.~\ref{fig:associahedron14}. Importantly, for a generic configuration of four two-dimensional vectors, the chamber fan always contains a chamber which is included in a single cone. This implies that the JK-residue computed in this chamber reduces to a single residue. However, there is no generalized triangle which covers the amplituhedron and therefore this cannot be the correct formula. Moreover, since our calculation can be related to the $i\epsilon$-prescription, we checked that the procedure outlined in \cite{Arkani-Hamed:2017tmz} also does not reproduce the correct volume function in this case. More generally, it does not work for odd values of $m$.


\section{Conclusions}
In this paper we have explored the application of the Jeffrey-Kirwan residue, a method widely used in supersymmetric localization calculations, to the problem of finding logarithmic differential forms for the tree amplituhedron. In particular, we have showed that  
the JK-residue provides the proper contour for the integrals \eqref{omegaintegral} with $k=1$, encoding the volume function for cyclic polytopes, and with $n-k-m_{even} = 1$ for the amplituhedra conjugate to even-dimensional cyclic polytopes. This contour does not rely on a \emph{a posteriori} analysis, i.e.~we do not need to use e.g.~the BCFW recursion relations to select the proper residues. Instead, we rely on the positivity of external data and follow the Jeffrey-Kirwan prescription. 
The computations we have performed allow us for an extensive study of the properties of cyclic polytopes and their conjugates. 
Our construction also provides a very systematic approach to find all regular triangulations of the amplituhedron. Therefore, it gives a plethora of equivalent representations of volume functions, connected to each other by the global residue theorem. All this is encoded in the beautiful and rich structure of the secondary polytope, which can be constructed by studying the chamber fan.

There are few natural questions which arise from our considerations. The most pressing one is whether the method developed in this paper can be generalized also for other amplituhedra: for higher helicity and beyond the tree level. The main obstacle is that, in these cases, the denominators in \eqref{omegaintegral} are not products of linear factors any more. In particular, the definition of charges is not a straightforward generalization of the cases we studied and, to our knowledge, there is no mathematical framework where such generalization has been explored. Our paper also poses new kind of questions which demand further systematic studies of the properties of amplituhedra. For example, we have only scratched the surface on understanding the parity conjugation in this context. Furthermore, for general amplituhedra, there is no classification of their possible triangulations. More specifically, one could consider the notion of {\em secondary amplituhedron}, an object which encapsulates all possible (regular) triangulations of the amplituhedron. Answering these questions might help us to prove many conjectural claims which have been made for the amplituhedron in the past years.


\section*{Acknowledgements}
We would like to thank Jacob Bourjaily, Mathew Bullimore and Sven Krippendorf for useful discussions. L.F. is supported by the Elitenetwork of Bavaria. This work was partially supported by the DFG Grant FE 1529/1-1.

\newpage

\bibliographystyle{nb}
\bibliography{JKresidue}

\end{document}